\documentclass[runningheads]{llncs}
\usepackage[T1]{fontenc}
\usepackage[caption=false]{subfig}
\usepackage{graphicx}
\usepackage{tabularx}

\usepackage{xcolor}
\definecolor{slate50}{HTML}{f8fafc}
\definecolor{slate100}{HTML}{f1f5f9}
\definecolor{slate200}{HTML}{e2e8f0}
\definecolor{slate300}{HTML}{cbd5e1}
\definecolor{slate400}{HTML}{94a3b8}
\definecolor{slate500}{HTML}{64748b}
\definecolor{slate600}{HTML}{475569}
\definecolor{slate700}{HTML}{334155}
\definecolor{slate800}{HTML}{1e293b}
\definecolor{slate900}{HTML}{0f172a}
\definecolor{slate950}{HTML}{020617}
\definecolor{gray50}{HTML}{f9fafb}
\definecolor{gray100}{HTML}{f3f4f6}
\definecolor{gray200}{HTML}{e5e7eb}
\definecolor{gray300}{HTML}{d1d5db}
\definecolor{gray400}{HTML}{9ca3af}
\definecolor{gray500}{HTML}{6b7280}
\definecolor{gray600}{HTML}{4b5563}
\definecolor{gray700}{HTML}{374151}
\definecolor{gray800}{HTML}{1f2937}
\definecolor{gray900}{HTML}{111827}
\definecolor{gray950}{HTML}{030712}
\definecolor{zinc50}{HTML}{fafafa}
\definecolor{zinc100}{HTML}{f4f4f5}
\definecolor{zinc200}{HTML}{e4e4e7}
\definecolor{zinc300}{HTML}{d4d4d8}
\definecolor{zinc400}{HTML}{a1a1aa}
\definecolor{zinc500}{HTML}{71717a}
\definecolor{zinc600}{HTML}{52525b}
\definecolor{zinc700}{HTML}{3f3f46}
\definecolor{zinc800}{HTML}{27272a}
\definecolor{zinc900}{HTML}{18181b}
\definecolor{zinc950}{HTML}{09090b}
\definecolor{neutral50}{HTML}{fafafa}
\definecolor{neutral100}{HTML}{f5f5f5}
\definecolor{neutral200}{HTML}{e5e5e5}
\definecolor{neutral300}{HTML}{d4d4d4}
\definecolor{neutral400}{HTML}{a3a3a3}
\definecolor{neutral500}{HTML}{737373}
\definecolor{neutral600}{HTML}{525252}
\definecolor{neutral700}{HTML}{404040}
\definecolor{neutral800}{HTML}{262626}
\definecolor{neutral900}{HTML}{171717}
\definecolor{neutral950}{HTML}{0a0a0a}
\definecolor{stone50}{HTML}{fafaf9}
\definecolor{stone100}{HTML}{f5f5f4}
\definecolor{stone200}{HTML}{e7e5e4}
\definecolor{stone300}{HTML}{d6d3d1}
\definecolor{stone400}{HTML}{a8a29e}
\definecolor{stone500}{HTML}{78716c}
\definecolor{stone600}{HTML}{57534e}
\definecolor{stone700}{HTML}{44403c}
\definecolor{stone800}{HTML}{292524}
\definecolor{stone900}{HTML}{1c1917}
\definecolor{stone950}{HTML}{0c0a09}
\definecolor{red50}{HTML}{fef2f2}
\definecolor{red100}{HTML}{fee2e2}
\definecolor{red200}{HTML}{fecaca}
\definecolor{red300}{HTML}{fca5a5}
\definecolor{red400}{HTML}{f87171}
\definecolor{red500}{HTML}{ef4444}
\definecolor{red600}{HTML}{dc2626}
\definecolor{red700}{HTML}{b91c1c}
\definecolor{red800}{HTML}{991b1b}
\definecolor{red900}{HTML}{7f1d1d}
\definecolor{red950}{HTML}{450a0a}
\definecolor{orange50}{HTML}{fff7ed}
\definecolor{orange100}{HTML}{ffedd5}
\definecolor{orange200}{HTML}{fed7aa}
\definecolor{orange300}{HTML}{fdba74}
\definecolor{orange400}{HTML}{fb923c}
\definecolor{orange500}{HTML}{f97316}
\definecolor{orange600}{HTML}{ea580c}
\definecolor{orange700}{HTML}{c2410c}
\definecolor{orange800}{HTML}{9a3412}
\definecolor{orange900}{HTML}{7c2d12}
\definecolor{orange950}{HTML}{431407}
\definecolor{amber50}{HTML}{fffbeb}
\definecolor{amber100}{HTML}{fef3c7}
\definecolor{amber200}{HTML}{fde68a}
\definecolor{amber300}{HTML}{fcd34d}
\definecolor{amber400}{HTML}{fbbf24}
\definecolor{amber500}{HTML}{f59e0b}
\definecolor{amber600}{HTML}{d97706}
\definecolor{amber700}{HTML}{b45309}
\definecolor{amber800}{HTML}{92400e}
\definecolor{amber900}{HTML}{78350f}
\definecolor{amber950}{HTML}{451a03}
\definecolor{yellow50}{HTML}{fefce8}
\definecolor{yellow100}{HTML}{fef9c3}
\definecolor{yellow200}{HTML}{fef08a}
\definecolor{yellow300}{HTML}{fde047}
\definecolor{yellow400}{HTML}{facc15}
\definecolor{yellow500}{HTML}{eab308}
\definecolor{yellow600}{HTML}{ca8a04}
\definecolor{yellow700}{HTML}{a16207}
\definecolor{yellow800}{HTML}{854d0e}
\definecolor{yellow900}{HTML}{713f12}
\definecolor{yellow950}{HTML}{422006}
\definecolor{lime50}{HTML}{f7fee7}
\definecolor{lime100}{HTML}{ecfccb}
\definecolor{lime200}{HTML}{d9f99d}
\definecolor{lime300}{HTML}{bef264}
\definecolor{lime400}{HTML}{a3e635}
\definecolor{lime500}{HTML}{84cc16}
\definecolor{lime600}{HTML}{65a30d}
\definecolor{lime700}{HTML}{4d7c0f}
\definecolor{lime800}{HTML}{3f6212}
\definecolor{lime900}{HTML}{365314}
\definecolor{lime950}{HTML}{1a2e05}
\definecolor{green50}{HTML}{f0fdf4}
\definecolor{green100}{HTML}{dcfce7}
\definecolor{green200}{HTML}{bbf7d0}
\definecolor{green300}{HTML}{86efac}
\definecolor{green400}{HTML}{4ade80}
\definecolor{green500}{HTML}{22c55e}
\definecolor{green600}{HTML}{16a34a}
\definecolor{green700}{HTML}{15803d}
\definecolor{green800}{HTML}{166534}
\definecolor{green900}{HTML}{14532d}
\definecolor{green950}{HTML}{052e16}
\definecolor{emerald50}{HTML}{ecfdf5}
\definecolor{emerald100}{HTML}{d1fae5}
\definecolor{emerald200}{HTML}{a7f3d0}
\definecolor{emerald300}{HTML}{6ee7b7}
\definecolor{emerald400}{HTML}{34d399}
\definecolor{emerald500}{HTML}{10b981}
\definecolor{emerald600}{HTML}{059669}
\definecolor{emerald700}{HTML}{047857}
\definecolor{emerald800}{HTML}{065f46}
\definecolor{emerald900}{HTML}{064e3b}
\definecolor{emerald950}{HTML}{022c22}
\definecolor{teal50}{HTML}{f0fdfa}
\definecolor{teal100}{HTML}{ccfbf1}
\definecolor{teal200}{HTML}{99f6e4}
\definecolor{teal300}{HTML}{5eead4}
\definecolor{teal400}{HTML}{2dd4bf}
\definecolor{teal500}{HTML}{14b8a6}
\definecolor{teal600}{HTML}{0d9488}
\definecolor{teal700}{HTML}{0f766e}
\definecolor{teal800}{HTML}{115e59}
\definecolor{teal900}{HTML}{134e4a}
\definecolor{teal950}{HTML}{042f2e}
\definecolor{cyan50}{HTML}{ecfeff}
\definecolor{cyan100}{HTML}{cffafe}
\definecolor{cyan200}{HTML}{a5f3fc}
\definecolor{cyan300}{HTML}{67e8f9}
\definecolor{cyan400}{HTML}{22d3ee}
\definecolor{cyan500}{HTML}{06b6d4}
\definecolor{cyan600}{HTML}{0891b2}
\definecolor{cyan700}{HTML}{0e7490}
\definecolor{cyan800}{HTML}{155e75}
\definecolor{cyan900}{HTML}{164e63}
\definecolor{cyan950}{HTML}{083344}
\definecolor{sky50}{HTML}{f0f9ff}
\definecolor{sky100}{HTML}{e0f2fe}
\definecolor{sky200}{HTML}{bae6fd}
\definecolor{sky300}{HTML}{7dd3fc}
\definecolor{sky400}{HTML}{38bdf8}
\definecolor{sky500}{HTML}{0ea5e9}
\definecolor{sky600}{HTML}{0284c7}
\definecolor{sky700}{HTML}{0369a1}
\definecolor{sky800}{HTML}{075985}
\definecolor{sky900}{HTML}{0c4a6e}
\definecolor{sky950}{HTML}{082f49}
\definecolor{blue50}{HTML}{eff6ff}
\definecolor{blue100}{HTML}{dbeafe}
\definecolor{blue200}{HTML}{bfdbfe}
\definecolor{blue300}{HTML}{93c5fd}
\definecolor{blue400}{HTML}{60a5fa}
\definecolor{blue500}{HTML}{3b82f6}
\definecolor{blue600}{HTML}{2563eb}
\definecolor{blue700}{HTML}{1d4ed8}
\definecolor{blue800}{HTML}{1e40af}
\definecolor{blue900}{HTML}{1e3a8a}
\definecolor{blue950}{HTML}{172554}
\definecolor{indigo50}{HTML}{eef2ff}
\definecolor{indigo100}{HTML}{e0e7ff}
\definecolor{indigo200}{HTML}{c7d2fe}
\definecolor{indigo300}{HTML}{a5b4fc}
\definecolor{indigo400}{HTML}{818cf8}
\definecolor{indigo500}{HTML}{6366f1}
\definecolor{indigo600}{HTML}{4f46e5}
\definecolor{indigo700}{HTML}{4338ca}
\definecolor{indigo800}{HTML}{3730a3}
\definecolor{indigo900}{HTML}{312e81}
\definecolor{indigo950}{HTML}{1e1b4b}
\definecolor{violet50}{HTML}{f5f3ff}
\definecolor{violet100}{HTML}{ede9fe}
\definecolor{violet200}{HTML}{ddd6fe}
\definecolor{violet300}{HTML}{c4b5fd}
\definecolor{violet400}{HTML}{a78bfa}
\definecolor{violet500}{HTML}{8b5cf6}
\definecolor{violet600}{HTML}{7c3aed}
\definecolor{violet700}{HTML}{6d28d9}
\definecolor{violet800}{HTML}{5b21b6}
\definecolor{violet900}{HTML}{4c1d95}
\definecolor{violet950}{HTML}{2e1065}
\definecolor{purple50}{HTML}{faf5ff}
\definecolor{purple100}{HTML}{f3e8ff}
\definecolor{purple200}{HTML}{e9d5ff}
\definecolor{purple300}{HTML}{d8b4fe}
\definecolor{purple400}{HTML}{c084fc}
\definecolor{purple500}{HTML}{a855f7}
\definecolor{purple600}{HTML}{9333ea}
\definecolor{purple700}{HTML}{7e22ce}
\definecolor{purple800}{HTML}{6b21a8}
\definecolor{purple900}{HTML}{581c87}
\definecolor{purple950}{HTML}{3b0764}
\definecolor{fuchsia50}{HTML}{fdf4ff}
\definecolor{fuchsia100}{HTML}{fae8ff}
\definecolor{fuchsia200}{HTML}{f5d0fe}
\definecolor{fuchsia300}{HTML}{f0abfc}
\definecolor{fuchsia400}{HTML}{e879f9}
\definecolor{fuchsia500}{HTML}{d946ef}
\definecolor{fuchsia600}{HTML}{c026d3}
\definecolor{fuchsia700}{HTML}{a21caf}
\definecolor{fuchsia800}{HTML}{86198f}
\definecolor{fuchsia900}{HTML}{701a75}
\definecolor{fuchsia950}{HTML}{4a044e}
\definecolor{pink50}{HTML}{fdf2f8}
\definecolor{pink100}{HTML}{fce7f3}
\definecolor{pink200}{HTML}{fbcfe8}
\definecolor{pink300}{HTML}{f9a8d4}
\definecolor{pink400}{HTML}{f472b6}
\definecolor{pink500}{HTML}{ec4899}
\definecolor{pink600}{HTML}{db2777}
\definecolor{pink700}{HTML}{be185d}
\definecolor{pink800}{HTML}{9d174d}
\definecolor{pink900}{HTML}{831843}
\definecolor{pink950}{HTML}{500724}
\definecolor{rose50}{HTML}{fff1f2}
\definecolor{rose100}{HTML}{ffe4e6}
\definecolor{rose200}{HTML}{fecdd3}
\definecolor{rose300}{HTML}{fda4af}
\definecolor{rose400}{HTML}{fb7185}
\definecolor{rose500}{HTML}{f43f5e}
\definecolor{rose600}{HTML}{e11d48}
\definecolor{rose700}{HTML}{be123c}
\definecolor{rose800}{HTML}{9f1239}
\definecolor{rose900}{HTML}{881337}
\definecolor{rose950}{HTML}{4c0519}

\usepackage{hyperref}
\usepackage[capitalize]{cleveref}
\crefformat{section}{\S#2#1#3}
\crefrangeformat{section}{\S#3#1#4 to \S#5#2#6}

\usepackage{xcolor}          

\hypersetup{
    colorlinks=true,         
    linkcolor=black,         
    citecolor=black,          
    urlcolor=black            
}

\usepackage{tikz}
\usepackage{pgfplots}
\usetikzlibrary{arrows.meta, backgrounds, calc, decorations.markings, fit, math, positioning, shapes.geometric}
\usetikzlibrary{pgfplots.statistics}

\usepackage{pgfplotstable}

\usepgfplotslibrary{fillbetween,groupplots}
\pgfplotsset{compat=1.15}

\pgfplotsset{keep if/.style args={#1 is #2}{
  x filter/.code={
    \edef\tempa{\thisrow{#1}}
    \edef\tempb{#2}
    \ifx\tempa\tempb
    \else
      
    \fi
  }
}}

\pgfplotsset{
    box plot legend/.style={
        legend image code/.code={%
          \draw[#1] (-1.5mm, -1mm) rectangle (1.5mm, 1mm);
          \draw[#1] (-0.2mm, -1mm) -- (-0.2mm, 1mm);
          \draw[#1] (-1.5mm, 0mm) -- (-3mm, 0mm) (-3mm, -1mm) -- (-3mm, 1mm);
          \draw[#1] (+1.5mm, 0mm) -- (+3mm, 0mm) (+3mm, -1mm) -- (+3mm, 1mm);
        }
    },
}

\usepackage{booktabs}
\usepackage{xspace}

\usepackage{listings}

\usepackage{multirow}
\usepackage{enumitem}

\makeatletter
\let\commentfullflexible\lst@column@fullflexible
\makeatother

\newcommand{\PFour}{$\mathtt{P4_{16}}$\xspace}

\def\BibTeX{{\rm B\kern-.05em{\sc i\kern-.025em b}\kern-.08em
    T\kern-.1667em\lower.7ex\hbox{E}\kern-.125emX}}

\let\oldparagraph\paragraph
\newcommand{\paragraphnodot}[1]{\oldparagraph{\texorpdfstring{\normalfont\textbf{#1}}{#1}}}
\renewcommand\paragraph[1]{\paragraphnodot{\texorpdfstring{\normalfont\textbf{#1.}}{#1}}}

\makeatletter
\newcommand\subparagraphh{\@startsection{subparagraph}{5}{\z@}%
  {3.25ex \@plus1ex \@minus .2ex}%
  {-1em}%
  {\normalfont\normalsize}{\raggedright}}
\makeatother

\let\oldsubparagraphh\subparagraphh
\newcommand{\subparagraphhnodot}[1]{\oldsubparagraphh{\texorpdfstring{\textit{\underline{#1}}}{#1}}}
\renewcommand\subparagraphh[1]{\subparagraphhnodot{\texorpdfstring{\textit{#1.}}{#1}}}

\begin{document}
\title{The Effects of iBGP Convergence}
%
%
\author{
  Roland Schmid \and
  Tibor Schneider \and
  Georgia Fragkouli \and
  Laurent Vanbever
}

\authorrunning{Schmid et al.}
%
\institute{ETH Zurich, Switzerland}
\maketitle              
\begin{abstract}
Analyzing violations of forwarding properties is a classic networking problem.
However, existing work is either tailored to the steady state---and not to \emph{transient} states during iBGP convergence---or does analyze transient violations but with inaccurate proxies, like control-plane convergence, or without precise control over the different impact factors.

We address this gap with a measurement framework that controllably and accurately measures transient violation times in realistic network deployments.
The framework relies on a programmable switch to flexibly emulate diverse topologies and gain traffic visibility at all links---enabling accurately inferring violation times of any forwarding property.
Using the framework, we analyze 50 network scenarios on a topology with 12 real routers, and show how factors like the network configuration and BGP event affect transient violation times.
Further, we shed light on less-known aspects of BGP convergence, including that transient violations can start \emph{before} the trigger event, or that keeping a backup route advertised at all times can \emph{increase} violation times.

\end{abstract}

\begin{figure*}[t]
  \centering
  \begin{tikzpicture}

  \newcommand{\BGPconvergence}{2.0035576} 

  \pgfplotsset{
    short legend/.style={%
      legend image code/.code={
        \draw[##1,line width=0.6pt]
          plot coordinates {
            (0cm,0cm)
            (0.3cm,0cm)
          };%
      }
    }
  }

  \begin{axis}[
    axis x line = center,
    axis y line = center,
    width=\linewidth,
    height=5cm,
    xlabel={violation},
    xlabel style={above},
    ylabel={\# routers\\recovered},
    y label style={anchor=south, text width=3cm, align=center},
    xmin=0,
    xmax=2.2,
    xtick={0.5,1,1.5,2},
    xticklabels={0.5\,s,1\,s,1.5\,s,2\,s},
    minor xtick={0.25,0.75,1.25,1.75},
    ymin=0,
    ymax=12,
    ytick={5, 10},
    yticklabels={5, 10},
    legend style={at={(0.43, 1.02)}, anchor=south, draw=none, /tikz/column 4/.style={column sep=5pt}, /tikz/column 2/.style={column sep=5pt}, inner sep=0pt},
    legend columns=3,
    clip mode=individual,
  ]

    \addlegendimage{short legend,thick,densely dashed,amber600}\addlegendentry{min prefix}
    \addlegendimage{short legend,thick,densely dotted,rose600}\addlegendentry{med prefix}
    \addlegendimage{short legend,thick,teal600}\addlegendentry{max prefix}

    \addplot[mark=none, draw=none, name path=yaxis] coordinates {
      (0,0)
      (0,11)
    };

    \addplot[mark=none, thick, draw=teal600, solid, name path=violation time 100.21.134.0/24] coordinates {
        (0,0)
        (0.792,1)
        (1.614,2)
        (1.661,3)
        (1.67,4)
        (1.679,5)
        (1.6840000000000002,6)
        (1.692,7)
        (1.715,8)
        (1.715,9)
        (1.733,10)
        (2.012,11)
    };
    \addplot[draw=none, fill=teal400, fill opacity=0.1] fill between[of=yaxis and violation time 100.21.134.0/24];

    \addplot[mark=none, thick, draw=rose600, densely dotted, name path=violation time 100.14.206.0/24] coordinates {
        (0,0)
        (0.28,1)
        (0.753,2)
        (0.8220000000000001,3)
        (0.8270000000000001,4)
        (0.8270000000000001,5)
        (0.838,6)
        (0.889,7)
        (1.0979999999999999,8)
        (1.125,9)
        (1.151,10)
        (1.151,11)
    };
    \addplot[draw=none, fill=rose400, fill opacity=0.1] fill between[of=yaxis and violation time 100.14.206.0/24];

    \addplot[mark=none, thick, draw=amber600, dashed, name path=violation time 100.11.240.0/24] coordinates {
        (0,0)
        (0.177,1)
        (0.212,2)
        (0.23800000000000002,3)
        (0.245,4)
        (0.245,5)
        (0.263,6)
        (0.27899999999999997,7)
        (0.35200000000000004,8)
        (0.35600000000000004,9)
        (0.597,10)
        (0.735,11)
    };
    \addplot[draw=none, fill=amber400, fill opacity=0.1] fill between[of=yaxis and violation time 100.11.240.0/24];

    \addplot[mark=none, very thick, solid, draw=black, clip=false] coordinates {
      (\BGPconvergence,0)
      (\BGPconvergence,12)
    } node [above, anchor=south, text width=3.5cm, align=center] {\it network-wide\\BGP convergence};

  \end{axis}
\end{tikzpicture}

  \vspace{-0.8cm}
  \caption{Transient violations differ across routers and prefixes. The first, median, and last prefix to recover reachability (out of ten randomly probed prefixes) after a withdraw event of 10k prefixes cannot be estimated from a network-wide BGP convergence time.}
  \label{fig:motivation}
\end{figure*}
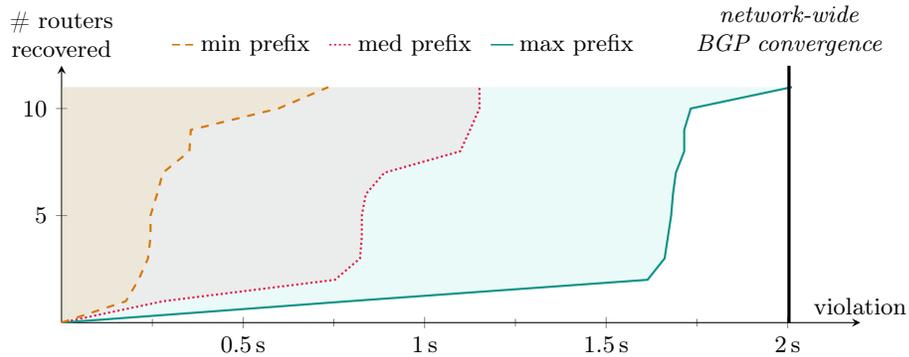

\section{Introduction}

Network verification has progressed a lot in the last decade, enabling operators to verify more complex (forwarding and routing) properties at scale, in networks with hundreds of devices.
Almost all existing network verifiers suffer from one key limitation though: they can only verify properties in the steady state, once the network has converged, not in transient states, while the network is converging.

Transient states are inherent to distributed routing protocols and can happen whenever the network undergoes an event such as a link failure or an external route change.
Concretely, this means that existing verifiers can assess a network to be formally correct--—for instance, assessing that traffic will \emph{never} bypass a firewall---even though the network can still be transiently incorrect---meaning that traffic can still transiently bypass the firewall.

Transient violations are especially problematic in the context of BGP, since the computation is per prefix and there can be \emph{many}.
Because of that, transient violations caused by BGP events can be particularly long. 
While transient violations induced by the convergence behavior of eBGP matter, it is beyond the control of a network operator and we focus on the \emph{effects of iBGP convergence}.

Simply ignoring transient states and violations is (unfortunately) not an option, for at least two reasons.
First, they happen frequently, as failures and external route changes are routine occurrences in large networks~\cite{bgp-instability-report}.
Second, they can potentially last for \emph{minutes} (cf.~\cref{sec:eval:processing}).

For operators though, accurately evaluating the impact of these transient violations on their network is challenging.
First, their actual duration---as our study shows---depends on many factors, including: the network topology, the configuration, the external routes received, and many others.
Second, transient violations can manifest themselves in short-lived episodes, possibly milliseconds long, but which repeat over a period.

Existing work on evaluating transient violations due to BGP (re-)convergence \cite{labovitz01theimpact,labovitz00delayed,park11quantifying,bush05happy,wang06ameasurementstudy,feldmann04measuring,schmid23predicting}, while useful, lacks in at least one of two ways.
First, work on convergence only considers control-plane messages, yielding convergence times that can be too pessimistic~\cite{bush05happy}.
Indeed, upon a multi-prefix event, the network-wide convergence time can be more than 2$\times$ the violation time for individual prefixes (cf.\ \cref{fig:motivation}).
Second, work that does consider transient violations either measures inaccurate proxies like single-router reaction times to BGP messages, or lacks precise control over the network configuration to test what-if scenarios and study their effects.

To sum up, the following research question remains open:
\emph{How can we measure and study transient violation times \emph{controllably} and \emph{accurately} during iBGP convergence triggered by BGP events?}

In this paper, we answer this question by presenting a complete measurement framework which flexibly emulates diverse network scenarios and precisely measures transient violation times in realistic network deployments.
There are two key challenges:
First, the framework must flexibly orchestrate real routers to emulate diverse physical network topologies and iBGP configurations.
To solve this challenge, the framework employs a programmable switch in combination with an automated centralized controller to orchestrate the network devices and emulating configurable physical link delays.

Second, the framework needs to accurately measure transient violations and enable pinpointing their root causes.
Specifically, accurately measuring transient violation times requires correctly identifying the converged states without affecting the convergence behavior, and ensuring that any software components can keep up with data-plane traffic used for measuring violations.
To solve this challenge, the framework correlates control- and data-plane traffic, and runs multiple instances of software components.
Further, accurately pinpointing the root causes of transient violations requires synchronized, full visibility of network traffic at each link, which is not a trivial task in any distributed system.
To this end, the framework configures the programmable switch to mirror all traffic to a centralized capturing device.

We have used the measurement framework to analyze 50 scenarios on the Abilene network topology, using 12 real routers.
Among others, our analysis of transient reachability violations shows that:
i)~transient violations can last up to minutes;
ii)~multi-prefix withdrawals cause the longest violations;
iii)~topological features like link propagation delays matter only for events involving few prefixes;
otherwise, the processing time dominates;
iv)~keeping backup routes advertised leads to lower or \emph{higher(!)} violation times depending on the BGP event.

Overall, we make the following \textbf{contributions}:
\begin{itemize}
  \item A controllable measurement framework (cf.\ \cref{sec:main}) along with orchestration code\footnote{We will make it publicly available after the submission.} for accurately measuring transient violations.
  \item An analysis of 50 network scenarios on the Abilene topology for transient reachability violations.
  \item A dataset of more than 4k pcaps recorded for various network scenarios on the Abilene topology with real routers. 
\end{itemize}

\section{Motivation}
\label{sec:motivation}

Consider the effects of a BGP withdraw event of 10k prefixes in an 11-router network showcased in \cref{fig:motivation}.
We observe a significant difference in violation times for the different prefixes affected by the event.
While some prefixes take the entire network-wide BGP convergence time observed, reachability for half of all prefixes is recovered in around half of that time.
Hence, we propose to analyze the BGP convergence with respect to the violations caused, instead of the mere time it takes to converge.
This section defines transient violations and uses a running example to showcase some characteristics of transient violations that makes measuring their duration challenging.

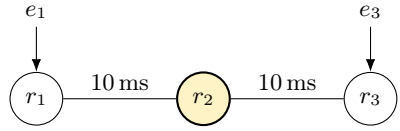
\begin{figure*}[t]
  \centering
  \begin{tikzpicture}[
  router/.style={circle, draw, inner sep=0pt, minimum size=7mm},
  ext/.style={draw=none},
  node distance=15mm,
]

  \node[router] (r1) {$r_{1}$};
  \node[router, right=of r1, thick, fill=amber100] (r2) {$r_{2}$};
  \node[router, right=of r2] (r3) {$r_{3}$};

  \node[ext, above=6mm of r1] (e1) {$e_1$};
  \node[ext, above=6mm of r3] (e3) {$e_3$};

  \draw (r1) -- node[pos=0.5, above] {10\,ms} (r2) -- node[pos=0.5, above] {10\,ms}(r3);
  \draw[->, >=Latex] (e1) -- (r1);
  \draw[->, >=Latex] (e3) -- (r3);

\end{tikzpicture}

  \caption{Example network with three routers $r_1$--$r_3$. Router $r_1$ and $r_{3}$ learn a route for prefix $p$ from its eBGP peer $e_1$ and $e_3$, respectively. Initially, all routers prefer $r_1$ as next hop. At time $t_0$, $e_1$ sends a withdraw message, causing the network to re-converge to using $r_3$ as next hop.}
  \label{fig:path3}
\end{figure*}

\paragraph{Running example}
Consider a ``path-3'' network that consists of three routers with link propagation delays as shown in \cref{fig:path3}.
The routers are configured in an iBGP full mesh.
For simplicity, all routers have the same control-plane processing delay of $x$\,ms, and negligible data-plane processing delay.
Router $r_1$ ($r_3$) learns a route for prefix $p$ from its eBGP peer $e_1$ ($e_3$).
Initially, the network is in a stable state where all the routers prefer the route through $r_1$, due to its shorter AS-path length.
At time $t_0$, $e_1$ sends a BGP withdraw message for $p$ to $r_1$.
This triggers the network re-converging to a new stable state, where all the routers prefer $r_3$ as the next hop.

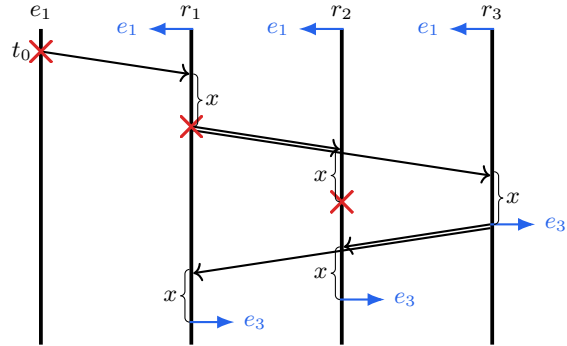
\begin{figure}[t]
  \centering
  \begin{tikzpicture}[
  yscale=-1,
  evaluate={
    real \e, \el, \er, \r, \rl, \rr, \barwidth;
    \barwidth = 0.05;
    \halfbarwidth = \barwidth * 0.5;
    \e1 = 0.0;
    \r1 = 2.0;
    \r2 = 4.0;
    \r3 = 6.0;
    \el1 = \e1 - \halfbarwidth;
    \er1 = \e1 + \halfbarwidth;
    \rl1 = \r1 - \halfbarwidth;
    \rr1 = \r1 + \halfbarwidth;
    \rl2 = \r2 - \halfbarwidth;
    \rr2 = \r2 + \halfbarwidth;
    \rl3 = \r3 - \halfbarwidth;
    \rr3 = \r3 + \halfbarwidth;
    real \x, \p, \t;
    \x = 0.7;
    \p = 0.3;
    int \i, \j;
    for \i in {0,...,10}{
      for \j in {0,...,10}{
        \t{\i,\j} = \i * \p + \j * \x;
      };
    };
    \t{start} = -0.3;
    \t{end} = \t{5,3} + 0.3;
  },
  drop/.style={minimum width=3mm, minimum height=3mm, inner sep=0pt, path picture={
    \draw[very thick, solid] (path picture bounding box.north east) -- (path picture bounding box.south west);
    \draw[very thick, solid] (path picture bounding box.north west) -- (path picture bounding box.south east);
  }},
  tick/.style={fill=black, rectangle, inner sep=0pt, minimum height=1.5pt, minimum width=5pt},
]
  \colorlet{leftColor}{blue600}
  \colorlet{rightColor}{blue600}
  \colorlet{dropColor}{red600}

  \newcommand{\GoesLeft}[3]{
    \draw[leftColor, thick, ->, >=Latex] ([xshift=\halfbarwidth * 1cm]#1, #2) -- ++(-6mm, 0) node[left] {$e_{1}$};
  }
  \newcommand{\GoesRight}[3]{
    \draw[rightColor, thick, ->, >=Latex] ([xshift=-\halfbarwidth * 1cm]#1, #2) -- ++(6mm, 0) node[right] {$e_{3}$};
  }
  \newcommand{\Drops}[3]{
    \node[drop, dropColor] at (#1, #2) {};
  }
  \newcommand{\NoteLeft}[4]{
    \node[coordinate] at (#1, #2) (point top) {};
    \node[coordinate] at (#1, #3) (point bot) {};
    \node[coordinate] at ($(point top)!0.5!(point bot)$) (point mid) {};
    \node[coordinate] at ($(point mid)-(0.6mm, 0)$) (point merge) {};
    \node[coordinate] at ($(point merge)-(0.6mm, 0)$) (point tip) {};
    \draw[rounded corners=0.6mm] (point top) -| (point merge) -- (point tip) node[left, inner sep=1pt] {#4};
    \draw[rounded corners=0.6mm] (point bot) -| (point merge) -- (point tip);
  }
  \newcommand{\NoteRight}[4]{
    \node[coordinate] at (#1, #2) (point top) {};
    \node[coordinate] at (#1, #3) (point bot) {};
    \node[coordinate] at ($(point top)!0.5!(point bot)$) (point mid) {};
    \node[coordinate] at ($(point mid)+(0.6mm, 0)$) (point merge) {};
    \node[coordinate] at ($(point merge)+(0.6mm, 0)$) (point tip) {};
    \draw[rounded corners=0.6mm] (point top) -| (point merge) -- (point tip) node[right, inner sep=1pt] {#4};
    \draw[rounded corners=0.6mm] (point bot) -| (point merge) -- (point tip);
  }

  \draw[line width=\barwidth * 1cm] (\e1,\t{start}) node[above] {$e_{1}$} -- (\e1, \t{end});
  \draw[line width=\barwidth * 1cm] (\r1,\t{start}) node[above] {$r_{1}$} -- (\r1, \t{end});
  \draw[line width=\barwidth * 1cm] (\r2,\t{start}) node[above] {$r_{2}$} -- (\r2, \t{end});
  \draw[line width=\barwidth * 1cm] (\r3,\t{start}) node[above] {$r_{3}$} -- (\r3, \t{end});

  \draw[thick, ->] (\e1, 0) node[left] {$t_0$} -- (\rl1, \t{1,0});
  \draw[thick, ->] (\rr1, \t{1,1}) -- (\rl2, \t{2,1});
  \draw[thick, ->, yshift=0.5mm] (\rr1, \t{1,1}) -- (\rl3, \t{3,1});
  \draw[thick, ->, yshift=0.5mm] (\rl3, \t{3,2}) -- (\rr1, \t{5,2});
  \draw[thick, ->] (\rl3, \t{3,2}) -- (\rr2, \t{4,2});
  
  \Drops{\e1}{\t{0,0}}{\t{0,0}}

  \GoesLeft{\r1}{\t{start}}{\t{1,1}}
  \Drops{\r1}{\t{1,1}}{\t{3,3}}
  \GoesRight{\r1}{\t{5,3}}{\t{end}}

  \GoesLeft{\r2}{\t{start}}{\t{2,2}}
  \Drops{\r2}{\t{2,2}}{\t{3,3}}
  \GoesRight{\r2}{\t{4,3}}{\t{end}}

  \GoesLeft{\r3}{\t{start}}{\t{3,2}}
  \GoesRight{\r3}{\t{3,2}}{\t{end}}

  \NoteRight{\rr1}{\t{1,0}}{\t{1,1}}{$x$}
  \NoteLeft{\rl1}{\t{5,2}}{\t{5,3}}{$x$}
  \NoteLeft{\rl2}{\t{2,1}}{\t{2,2}}{$x$}
  \NoteRight{\rr3}{\t{3,1}}{\t{3,2}}{$x$}
  \NoteLeft{\rl2}{\t{4,2}}{\t{4,3}}{$x$}
\end{tikzpicture}


  \caption{The sequence diagram of messages and forwarding states that the withdrawal event triggers in the network of \cref{fig:path3}.}
  \label{fig:simple_violation_sequence}
\end{figure}

\subparagraphh{Routing events cause transient violations}
Consider the sequence diagram of messages and forwarding states triggered by the withdrawal (cf.\ \cref{fig:simple_violation_sequence}).
The withdrawal causes some routers to temporarily drop traffic, while the routers have no route to the prefix (referred to as a \emph{black hole}).
Specifically, once $r_1$ has received and processed the withdrawal, it installs a black hole and sends withdraw messages to $r_2$ and $r_3$.
Upon receiving/processing the withdrawal, $r_2$ installs a black hole, while $r_3$ shifts to using its local backup route and announces it to $r_1$ and $r_2$.
Eventually, $r_1$ and $r_2$ receive/process the announcements, thus shifting to their final forwarding state where they forward to the right.

\begin{definition}[transiently-violating packet]
  A packet transiently violates reachability if i)~the packet is dropped
  due to a BGP event, but ii)~the packet would not violate reachability in the stables state before or after the BGP re-convergence triggered by the event.
\end{definition}
An example of a violating packet is one that originates at $r_2$ while $r_2$ has a black hole, hence $r_2$ drops it.
In contrast, a non-violating packet is one that originates at $r_2$ after $r_2$ has started forwarding to the right.
The packet eventually reaches $r_3$ and successfully gets forwarded through $e_3$.

However, violating packets are not only the ones that originate at routers during black holes.
Even worse, a violating packet can originate at a router \emph{before} the BGP event happens!
In our running example, consider a packet leaving $r_2$ at $t_0-20$\,ms: at this time, the network is in a stable state where all routers forward through $e_1$.
By the time the packet reaches $e_1$, at $t_0$, the withdrawal has already happened.
So $e_1$ drops the packet, which transiently violates reachability.
In contrast, a packet leaving $r_2$ just before $t_0-20$\,ms reaches $e_1$ just before the withdrawal, hence it is a non-violating one.
\begin{definition}[transient violation time]
The transient violation time at a router $r$ toward prefix $p$ is the duration during which any packet sent at router~$r$ to the prefix~$p$ would transiently violate reachability.
\end{definition}
This definition captures worst-case violation times.
It is agnostic of the actual traffic sent at a router, and instead considers the entire duration during which any hypothetical traffic sent at a router would violate reachability.
In our example, $r_2$ transiently violates reachability for $3x+60$\,ms:
As discussed above, the violation starts before the withdrawal, at $t_0-20$\,ms, as in-flight traffic leaving $r_2$ at this time is eventually dropped at $e_1$.
The violation continues all the way to $t_0+3x+40$\,ms, when $r_2$ shifts to using $e_3$.

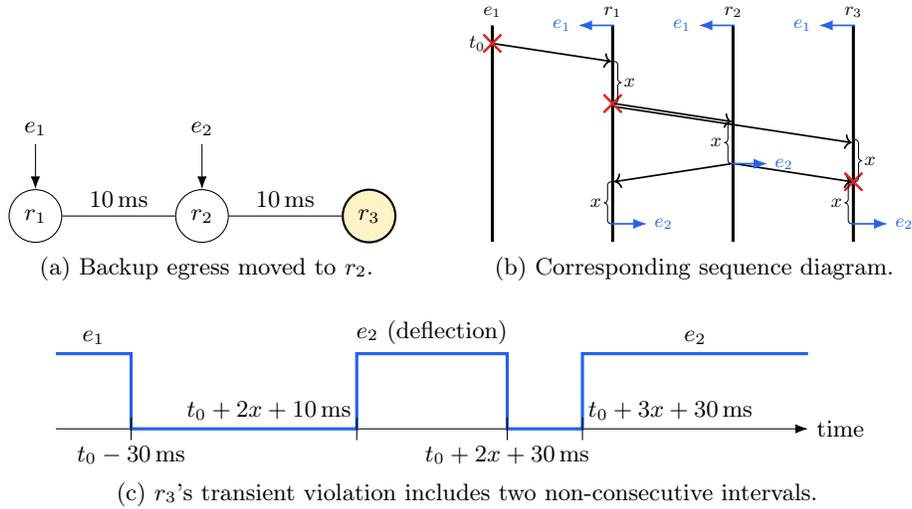
\begin{figure}[t]
  \centering
  \subfloat[Backup egress moved to $r_2$.\label{fig:discontinuous_violation_network}]{%
    \begin{tikzpicture}[
  router/.style={circle, draw, inner sep=0pt, minimum size=7mm},
  ext/.style={draw=none},
  node distance=15mm,
]

  \node[router] (r1) {$r_{1}$};
  \node[router, right=of r1] (r2) {$r_{2}$};
  \node[router, right=of r2, thick, fill=amber100] (r3) {$r_{3}$};

  \node[ext, above=6mm of r1] (e1) {$e_1$};
  \node[ext, above=6mm of r2] (e2) {$e_2$};

  \draw (r1) -- node[pos=0.5, above] {10\,ms} (r2) -- node[pos=0.5, above] {10\,ms}(r3);
  \draw[->, >=Latex] (e1) -- (r1);
  \draw[->, >=Latex] (e2) -- (r2);

\end{tikzpicture}

%
  }
  \hfill
  \subfloat[Corresponding sequence diagram.\label{fig:discontinuous_violation_sequence}]{%
    \scalebox{0.8}{%
      \begin{tikzpicture}[
  yscale=-1,
  evaluate={
    real \e, \el, \er, \r, \rl, \rr, \barwidth;
    \barwidth = 0.05;
    \halfbarwidth = \barwidth * 0.5;
    \e1 = 0.0;
    \r1 = 2.0;
    \r2 = 4.0;
    \r3 = 6.0;
    \el1 = \e1 - \halfbarwidth;
    \er1 = \e1 + \halfbarwidth;
    \rl1 = \r1 - \halfbarwidth;
    \rr1 = \r1 + \halfbarwidth;
    \rl2 = \r2 - \halfbarwidth;
    \rr2 = \r2 + \halfbarwidth;
    \rl3 = \r3 - \halfbarwidth;
    \rr3 = \r3 + \halfbarwidth;
    real \x, \p, \t;
    \x = 0.7;
    \p = 0.3;
    int \i, \j;
    for \i in {0,...,10}{
      for \j in {0,...,10}{
        \t{\i,\j} = \i * \p + \j * \x;
      };
    };
    \t{start} = -0.3;
    \t{end} = \t{3,3} + 0.3;
  },
  drop/.style={minimum width=3mm, minimum height=3mm, inner sep=0pt, path picture={
    \draw[very thick, solid] (path picture bounding box.north east) -- (path picture bounding box.south west);
    \draw[very thick, solid] (path picture bounding box.north west) -- (path picture bounding box.south east);
  }},
  tick/.style={fill=black, rectangle, inner sep=0pt, minimum height=1.5pt, minimum width=5pt},
]
  \colorlet{leftColor}{blue600}
  \colorlet{rightColor}{blue600}
  \colorlet{dropColor}{red600}

  \newcommand{\GoesLeft}[3]{
    \draw[leftColor, thick, ->, >=Latex] ([xshift=\halfbarwidth * 1cm]#1, #2) -- ++(-6mm, 0) node[left] {$e_{1}$};
  }
  \newcommand{\GoesRight}[3]{
    \draw[rightColor, thick, ->, >=Latex] ([xshift=-\halfbarwidth * 1cm]#1, #2) -- ++(6mm, 0) node[right] {$e_{2}$};
  }
  \newcommand{\Drops}[3]{
    \node[drop, dropColor] at (#1, #2) {};
  }
  \newcommand{\NoteLeft}[4]{
    \node[coordinate] at (#1, #2) (point top) {};
    \node[coordinate] at (#1, #3) (point bot) {};
    \node[coordinate] at ($(point top)!0.5!(point bot)$) (point mid) {};
    \node[coordinate] at ($(point mid)-(0.6mm, 0)$) (point merge) {};
    \node[coordinate] at ($(point merge)-(0.6mm, 0)$) (point tip) {};
    \draw[rounded corners=0.6mm] (point top) -| (point merge) -- (point tip) node[left, inner sep=1pt] {#4};
    \draw[rounded corners=0.6mm] (point bot) -| (point merge) -- (point tip);
  }
  \newcommand{\NoteRight}[4]{
    \node[coordinate] at (#1, #2) (point top) {};
    \node[coordinate] at (#1, #3) (point bot) {};
    \node[coordinate] at ($(point top)!0.5!(point bot)$) (point mid) {};
    \node[coordinate] at ($(point mid)+(0.6mm, 0)$) (point merge) {};
    \node[coordinate] at ($(point merge)+(0.6mm, 0)$) (point tip) {};
    \draw[rounded corners=0.6mm] (point top) -| (point merge) -- (point tip) node[right, inner sep=1pt] {#4};
    \draw[rounded corners=0.6mm] (point bot) -| (point merge) -- (point tip);
  }

  \draw[line width=\barwidth * 1cm] (\e1,\t{start}) node[above] {$e_{1}$} -- (\e1, \t{end});
  \draw[line width=\barwidth * 1cm] (\r1,\t{start}) node[above] {$r_{1}$} -- (\r1, \t{end});
  \draw[line width=\barwidth * 1cm] (\r2,\t{start}) node[above] {$r_{2}$} -- (\r2, \t{end});
  \draw[line width=\barwidth * 1cm] (\r3,\t{start}) node[above] {$r_{3}$} -- (\r3, \t{end});

  \draw[thick, ->] (\e1, 0) node[left] {$t_0$} -- (\rl1, \t{1,0});
  \draw[thick, ->] (\rr1, \t{1,1}) -- (\rl2, \t{2,1});
  \draw[thick, ->, yshift=0.5mm] (\rr1, \t{1,1}) -- (\rl3, \t{3,1});
  \draw[thick, ->] (\rl2, \t{2,2}) -- (\rr1, \t{3,2});
  \draw[thick, ->] (\rr2, \t{2,2}) -- (\rl3, \t{3,2});
  
  \Drops{\e1}{\t{0,0}}{\t{0,0}}

  \GoesLeft{\r1}{\t{start}}{\t{1,1}}
  \Drops{\r1}{\t{1,1}}{\t{3,3}}
  \GoesRight{\r1}{\t{3,3}}{\t{end}}

  \GoesLeft{\r2}{\t{start}}{\t{2,2}}
  \GoesRight{\r2}{\t{2,2}}{\t{end}}

  \GoesLeft{\r3}{\t{start}}{\t{3,2}}
  \Drops{\r3}{\t{3,2}}{\t{3,3}}
  \GoesRight{\r3}{\t{3,3}}{\t{end}}

  \NoteRight{\rr1}{\t{1,0}}{\t{1,1}}{$x$}
  \NoteLeft{\rl1}{\t{3,2}}{\t{3,3}}{$x$}
  \NoteLeft{\rl2}{\t{2,1}}{\t{2,2}}{$x$}
  \NoteRight{\rr3}{\t{3,1}}{\t{3,2}}{$x$}
  \NoteLeft{\rl3}{\t{3,2}}{\t{3,3}}{$x$}
\end{tikzpicture}

%
    }%
  }
  \\
  \subfloat[$r_3$'s transient violation includes two non-consecutive intervals.\label{fig:discontinuous_violation_intervals}]{%
    \begin{tikzpicture}[
  evaluate={
    real \drop, \e;
    \drop = 0.0;
    \e1 = 1.0;
    \e2 = 1.0;
  },
]

  \newcommand{\Tick}[3]{
    \draw (#1, 0.15) -- (#1, -0.15) node[inner sep=2pt, #3]{#2};
  }

  \draw[->, >=Latex] (0, 0) -- (10, 0) node[right] {time};
  \draw[very thick, blue600] (0, \e1) -- (1, \e1) node[pos=0.5, above, black] {$e_{1}$} 
  -- (1, \drop) -- (4, \drop)
  -- (4, \e2) -- (6, \e2) node[pos=0.5, above, black] {$e_2$ (deflection)}
  -- (6, \drop) -- (7, \drop)
  -- (7, \e2) -- (10, \e2) node[pos=0.5, above, black] {$e_2$};

  \Tick{1}{$t_0 - 30$\,ms}{below}
  \Tick{4}{$t_0 + 2x + 10$\,ms}{above left, yshift=2mm}
  \Tick{6}{$t_0 + 2x + 30$\,ms}{below}
  \Tick{7}{$t_0 + 3x + 30$\,ms}{above right, yshift=2mm}

\end{tikzpicture}

%
  }
  \caption{Example network where $r_3$ experiences non-consecutive transient violations. The sequence diagram of messages and forwarding states that the withdrawal event triggers in the network of \cref{fig:discontinuous_violation_network}.}
  \label{fig:discontinuous_violation}
\end{figure}

\subparagraphh{Different routers can experience violations at different times}
In our example, the reachability violation at $r_1$ starts at $t_0-10$\,ms, i.e., after it starts for $r_2$.
This is because $r_1$ is closer to $e_1$, so it learns earlier about the withdrawal, and has for a shorter interval in-flight traffic dropped at $e_1$.

\subparagraphh{Transient violations can be non-consecutive}
To showcase another interesting characteristic of transient violations, let us slightly modify the network topology to the one of \cref{fig:discontinuous_violation_network}.
Here, instead of $r_3$, $r_2$ learns a backup route from $e_2$ for prefix $p$.
The initial and final forwading states do not change (cf.\ \cref{fig:discontinuous_violation_sequence}).

What changes, though, is that traffic from $r_3$ temporarily gets deflected.
This splits the transient violation time into two non-consecutive intervals (cf.\ \cref{fig:discontinuous_violation_intervals}): i)~($t_0-30$\,ms, $t_0+2x+10$\,ms), where $e_3$'s traffic is dropped at $e_1$, as $r_3$ has not yet learned about the withdrawal and its traffic is not yet deflected through $e_2$, and ii)~($t_0+2x+30$\,ms, $t_0+3x+30$\,ms), where $e_3$ drops the traffic due to a black hole, as it has processed $e_1$'s withdrawal but not $e_2$'s announcement.

\section{Measuring Transient Violation Times}
\label{sec:main}
We design a framework for studying transient violations for varying the network topology and configuration.
The framework must trigger BGP events and accurately measure the resulting transient forwarding anomalies at a router level for the entire network \emph{without} disturbing the control plane and thus interferring with the measurements.

\paragraph{Framework design}
Our framework allows to study transient violations caused by BGP (re-)convergence \emph{controllably} and \emph{accurately} in a 2-step procedure:
\begin{enumerate}
  \item We design a measurement framework (cf.\ \cref{sec:main:testbed}) that orchestrates real routers along with software components to emulate a network topology, inject data-plane traffic, and introduce a routing event.
  This allows us to capture the network-wide effects of iBGP convergence with a given network configuration by collecting all of the network's traffic as one \emph{sample}.
  \item We analyze the collected samples offline (cf.\ \cref{sec:main:pipeline}) to infer the transient violations, at each router in the network.
\end{enumerate}

\subsection{Measurement framework}
\label{sec:main:testbed}

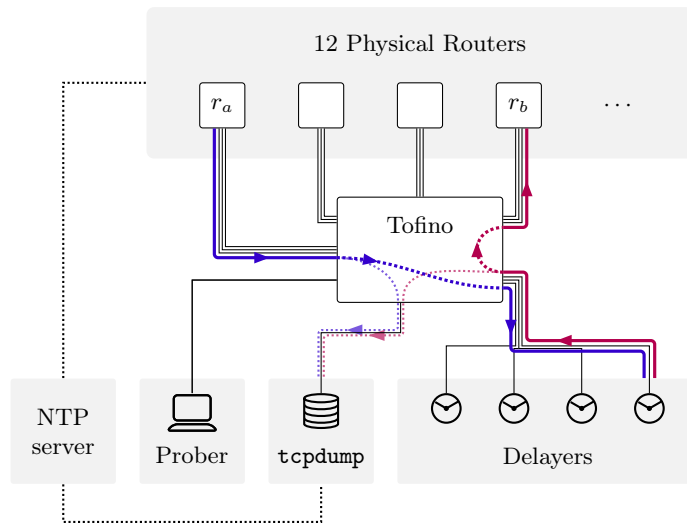
\begin{figure}[t]
  \centering
  \def\ShowProberTraffic{1}
  \colorlet{NotDelayed}{red!20!blue}
\colorlet{IsDelayed}{red!70!blue}
\begin{tikzpicture} [
    router/.style={rounded corners=0.4mm, draw, fill=white, minimum size=6mm},
    triple/.style={draw, preaction={draw, double, double distance=2pt}},
    box/.style={draw, rounded corners=0.4mm},
    component/.style={box, draw=none, fill=black!5, inner sep=0pt, minimum height=14mm},
    database/.style={cylinder, draw, shape border rotate=90, aspect=0.5, path picture={
        \coordinate (w1) at ($(path picture bounding box.north west)!0.3!(path picture bounding box.south west)$);
        \coordinate (e1) at ($(path picture bounding box.north east)!0.3!(path picture bounding box.south east)$);
        \coordinate (w2) at ($(path picture bounding box.north west)!0.5!(path picture bounding box.south west)$);
        \coordinate (e2) at ($(path picture bounding box.north east)!0.5!(path picture bounding box.south east)$);
        \coordinate (w3) at ($(path picture bounding box.north west)!0.7!(path picture bounding box.south west)$);
        \coordinate (e3) at ($(path picture bounding box.north east)!0.7!(path picture bounding box.south east)$);
        \draw (w1) to[out=-90, in=-90, looseness=0.5] (e1);
        \draw (w2) to[out=-90, in=-90, looseness=0.5] (e2);
        \draw (w3) to[out=-90, in=-90, looseness=0.5] (e3);
    }},
    clock/.style={circle, draw, path picture={
      \coordinate (c) at (path picture bounding box.center);
      \draw[fill] (c) circle (0.2mm);
      \draw (c) --++ (150:1.7mm);
      \draw (c) --++ (30:2.3mm);
    }},
    laptop/.style={path picture={
      \coordinate[yshift=-0.4pt] (screen nw) at ($(path picture bounding box.north west)!0.15!(path picture bounding box.north east)$);
      \coordinate (screen height w) at ($(path picture bounding box.north west)!0.7!(path picture bounding box.south west)$);
      \coordinate (screen height e) at ($(path picture bounding box.north east)!0.7!(path picture bounding box.south east)$);
      \coordinate (screen se) at ($(screen height w)!0.85!(screen height e)$);
      \coordinate[yshift=-0.5mm] (keys nw) at ($(screen nw |- screen se)$);
      \coordinate[yshift=-0.5mm] (keys ne) at (screen se);
      \coordinate[yshift=+0.4pt, xshift=+0.4pt] (keys sw) at (path picture bounding box.south west);
      \coordinate[yshift=+0.4pt, xshift=-0.4pt] (keys se) at (path picture bounding box.south east);
      \draw[rounded corners=0.5mm] (screen nw) rectangle (screen se);
      \draw[rounded corners=0.5mm] ($(keys nw)!0.5!(keys ne)$) -- (keys nw) -- (keys sw) -- (keys se) -- (keys ne) -- cycle;
    }},
    drop/.style={minimum width=2mm, minimum height=2mm, inner sep=0pt, path picture={
      \draw[very thick, draw=red, solid] (path picture bounding box.north east) -- (path picture bounding box.south west);
      \draw[very thick, draw=red, solid] (path picture bounding box.north west) -- (path picture bounding box.south east);
    }},
    mid arrow/.style={postaction={decorate,decoration={
          markings,
          mark=at position #1 with {\arrow{Triangle[length=6pt, width=4pt, sep=-3.2pt]}}
    }}},
    iface/.style={coordinate},
  ]

  \node[router] (r1) {$r_{a}$};
  \node[router, right=7mm of r1] (r2) {};
  \node[router, right=7mm of r2] (r3) {};
  \node[router, right=7mm of r3] (r4) {$r_{b}$};
  \node[router, right=7mm of r4, opacity=0, label=center:$\cdots$] (rn) {};
  \begin{pgfonlayer}{background}
    \node[rounded corners=1mm, fit={(r1) (r2) (r3) (r4) (rn)}, draw=none, fill=black!5, inner sep=7mm, yshift=3mm] (physical) {};
  \end{pgfonlayer}
  \node[yshift=5mm] at (physical) {12 Physical Routers};

  \node[box, minimum width=22mm, minimum height=14mm, below=5mm of physical, label={[yshift=-6mm]Tofino}] (tofino) {};

  \draw[triple] (r1) |- (tofino.180) node[coordinate] (tofino if1) {};
  \draw[triple] (r2) |- (tofino.160) node[coordinate] (tofino if2) {};
  \draw[triple] (r3) -- (tofino.north) node[coordinate] (tofino if3) {};
  \draw[triple] (r4) |- (tofino.20) node[coordinate] (tofino if4) {};

  \node[component, minimum width=4cm, below=10mm of tofino, xshift=17mm] (delayer) {};
  \node[iface, xshift=-1.35cm] at (delayer.north) (delayer if1) {};
  \node[iface, xshift=-0.45cm] at (delayer.north) (delayer if2) {};
  \node[iface, xshift=+0.45cm] at (delayer.north) (delayer if3) {};
  \node[iface, xshift=+1.35cm] at (delayer.north) (delayer if4) {};
  \node[clock, thick, minimum size=3.8mm, below=2mm of delayer if1] (delayer 1) {};
  \node[clock, thick, minimum size=3.8mm, below=2mm of delayer if2] (delayer 2) {};
  \node[clock, thick, minimum size=3.8mm, below=2mm of delayer if3] (delayer 3) {};
  \node[clock, thick, minimum size=3.8mm, below=2mm of delayer if4] (delayer 4) {};
  \node[above=1mm of delayer.south] {Delayers};
  \node[coordinate] at ($(r4)-(0, 3.2)$) (delayer p1) {};
  \node[coordinate, yshift=-1pt] at (delayer p1) (delayer p2) {};
  \draw[shorten >=1pt] (delayer 1) |- (delayer p1);
  \draw[] (delayer 2) |- (delayer p2);
  \draw[] (delayer 3) |- (delayer p2);
  \draw[shorten >=1pt] (delayer 4) |- (delayer p1);
  \draw[] (delayer p2) -- (delayer p1);
  \draw[triple] (delayer p1) |- (tofino.340) node[coordinate] (tofino if6) {};

  \node[component, minimum width=1.4cm, left=3mm of delayer] (collector) {};
  \node[above=1mm of collector.south, align=center] (collector text) {\texttt{tcpdump}};
  \node[iface] at (collector.north) (collector if) {};
  \node[database, thick, minimum width=4.5mm, minimum height=4.5mm, below=2mm of collector if] (collector db) {};
  \draw[] (collector db) -- ++(0, 11mm) node[coordinate] (collector p1) {} -| (tofino.250) node[coordinate] (tofino if7) {};

  \node[component, minimum width=1.4cm, left=3mm of collector] (prober) {};
  \node[base left=4.2mm of collector text, align=center] {Prober};
  \node[iface] at (prober.north) (prober if) {};
  \node[laptop, thick, minimum width=7mm, minimum height=5mm, below=2mm of prober if] (prober laptop) {};
  \draw[] (prober laptop) |- (tofino.200) node[coordinate] (tofino if8) {};

  \node[component, minimum width=1.4cm, left=3mm of prober] (ntp) {};
  \node[align=center] at (ntp) {NTP\\server};
  \node[iface] at (ntp.north) (ntp if) {};
  \node[laptop, thick, minimum width=7mm, minimum height=5mm, below=2mm of prober if] (prober laptop) {};
  \draw[] (prober laptop) |- (tofino.200) node[coordinate] (tofino if8) {};

  \draw[thick, densely dotted] (ntp if) |- (physical) (ntp.south) -- ++(0, -5mm) -| (collector.south);

  \draw[mid arrow=0.7, very thick, NotDelayed, rounded corners=0.5mm] ([xshift=-3pt]r1.south) |- ([yshift=-3pt]tofino if1) node[coordinate] (path p1) {};
  \draw[mid arrow=0.2, very thick, NotDelayed, rounded corners=0.5mm, densely dotted] (path p1) to[out=0, in=180] ([yshift=-3pt]tofino if6) node[coordinate] (path p2) {};
  \draw[mid arrow=0.2, very thick, NotDelayed, rounded corners=0.5mm] (path p2) -| ([xshift=-3pt, yshift=-2pt] delayer p1) -- ([xshift=-2pt,yshift=-2pt]delayer p1 -| delayer if4) -- ([xshift=-2pt]delayer if4);

  \draw[mid arrow=0.5, very thick, IsDelayed,  rounded corners=0.5mm] ([xshift=2pt]delayer if4) |- ([xshift=3pt, yshift=3pt]delayer p2) |- ([yshift=3pt]tofino if6) node[coordinate] (path p3) {};
  \draw[mid arrow=0.5, very thick, IsDelayed,  rounded corners=0.5mm, densely dotted] (path p3) to[out=180, in=180, looseness=2] ([yshift=-3pt]tofino if4) node[coordinate] (path p4) {};
  \draw[mid arrow=0.5, very thick, IsDelayed,  rounded corners=0.5mm] (path p4) -| ([xshift=3pt]r4.south);

  \begin{pgfonlayer}{background}
    \draw[mid arrow=0.65, thick, NotDelayed!65, rounded corners=0.5mm, densely dotted] (path p1) to[out=0, in=90, looseness=1] ([xshift=-1pt]tofino if7) |- ([xshift=-1pt, yshift=1pt]collector p1) -- ([xshift=-1pt]collector if);
    \draw[mid arrow=0.65, thick, IsDelayed!65,  rounded corners=0.5mm, densely dotted] (path p3) to[out=180, in=90, looseness=1] ([xshift=+1pt]tofino if7) |- ([xshift=1pt, yshift=-1pt]collector p1) -- ([xshift=+1pt]collector if);
  \end{pgfonlayer}

\end{tikzpicture}

  \caption{
    Illustration of the testbed architecture.
    The thick blue and red lines shows the path of a single packet from $r_{a}$ to $r_{b}$ that is delayed and mirrored for the offline analysis.
  }
  \label{fig:testbed}
\end{figure}

This section details the design of our measurement framework (cf.~\cref{fig:testbed}), which employs a programmable switch to connect all the components.
This enables both easy setup, as all devices simply connect to the programmable switch, and gives us the flexibility to \emph{(i)}~automatically change the topology by reprogramming the switch, and \emph{(ii)}~intercept packets on all links of the network and mirror them to a capturing device.

\paragraph{Emulating a physical topology}
Our testbed allows to configure connections between routers and their distance by emulating propagation delay.
The physical connections can be rewired via static configuration of the Intel Tofino programmable switch (Tofino), which forwards all packets from a given ingress port to the corresponding egress port at negligible latency.
Adding configurable propagation delay, however, requires keeping packets in a queue for multiple milliseconds which can be challenging to achieve for a programmable switch at high bandwidth.
Therefore, we develop a custom \emph{Delayer} program that expects packets with a custom packet header that specifies the delay to apply in µs, and sends all packets back after the given duration.
To apply a custom link propagation delay, the Tofino adds the custom headers specified for each ingress-egress-pair, stores additional routing information in the custom header to guarantee that packets always take the correct egress and are not really \emph{routed} by the programmable switch, and sent off to the delayer.
When receiving the delayed packets from the Delayer, the Tofino removes the custom header and forwards the packets according to the routing information stored in the custom header.

For some scenarios, especially those with large propagation delays and a lot of traffic, we noticed that a single Delayer cannot keep up with the demand.
We therefore run four Delayer instances on separate interfaces, and implement a round-robin based load balancer directly on the programmable switch.

\paragraph{Configuring the network devices}
We configure all routers with both OSPF and BGP, using an abstraction provided by the open-source BGP simulator BGPsim~\cite{schneider2021snowcap}.
While originally designed as a network control-plane simulator, it also allows to export configurations for Cisco- and FRR-compatible hardware devices.
We extend the simulator with our own orchestration library to manage and maintain \texttt{ssh} connections with the routers and deploying the configurations.

\paragraph{Emulating BGP routing inputs}
The framework configures and launches an ExaBGP instance on a server connected to the programmable switch, connecting to each border router on a dedicated interface.
This allows us to quickly setup different routing inputs, which we tested for up to one million prefixes per session.

When triggering large BGP events, however, we observe that ExaBGP generates BGP messages slower than physical routers would.
In order to emulate realistic scenarios seen in the Internet, we thus reserve one physical router to trigger the external BGP events by modifying a single route map.

\paragraph{Establishing a converged state}
In order to gain controllability over transient violation experiments, it is critical that we perform the event on a correctly configured, converged, initial state, and measure the entire convergence process.
The initial state must exactly match the expectation to ensure a repeatable setup and avoid erroneous states, possibly caused by an eariler experiment.
Conversely, we must ensure that the event we measure has fully converged without disturbing the convergence process itself.
To that end, we use two separate approaches to verify the converged state of the network before and during/after the event.

Before the experiment, we establish that both OSPF and BGP are in the correct state.
For OSPF, we check the next-hop and cost from each router to each other router is as expected.
For BGP, we first wait until all BGP sessions are established, and then wait until all routers learn their expected BGP routes.
We do so by periodically issuing \texttt{show} commands via the CLI and parsing its output automatically.

Using a similar approach for checking the triggered event has fully converged will put additional load on the routers' BGP processes, thus affecting the convergence behavior.
Instead, we check for convergence by relying on the mirrored traffic; we assume that a network has converged if we do not observe any BGP update messages for ten seconds (ignoring keep-alive messages).
Notice that this technique does not absolutely guarantee convergence.
It is theoretically possible that it takes over ten seconds for a router to process a single BGP message.
For our experiments, we thus employ an additional verification step to ensure this never occurs (cf.\ \cref{sec:main:pipeline}).

\paragraph{Measuring the network state}
We rely on data-plane traffic probes to observe which router can reach which prefix, and how that changes over time.
To that end, our framework generates probe traffic from each source router to a subset of destinations at a constant rate, adding a sequence number to the packets' payload.
The programmable switch then mirrors all traffic (both prober traffic and control-plane messages) to a server to capture the packets and store them for further processing.
Note that we mirror traffic both \emph{before} and \emph{after} passing through the Delayer (cf.\ \cref{fig:testbed}).
In other words, we capture traffic at both ends of each link in the network to obtain a unique synchronized view of the network.

\paragraph{Orchestrating the testbed}
We use a centralized controller to automatically orchestrate the testbed and measure convergence events.
It performs the following steps to collect one sample for a given scenario:
\begin{enumerate}
  \item The controller configures the Tofino and all routers according to the scenario.
  \item It configures and starts an ExaBGP instance to inject external BGP routes, and waits until the network reaches the expected initial converged state.
  \item Next, the controller starts the prober traffic and starts capturing all mirrored traffic on the server into a single pcap file for further processing.
  \item Five seconds later, the controller triggers the network event.
  \item As soon as the network has converged, the controller stops all processes, compresses the collected data and stores the pcap file for further processing.
\end{enumerate}

\paragraph{Implementation}
Overall, we wrote 16k lines of Rust code to implement the software components of our measurement framework (delayers, prober, and controller).
In addition, we wrote 800 lines of \PFour code for the data plane of the programmable switch and 500 lines of python for configuring its control-plane, allowing to flexibly emulate different network topologies in our testbed.\footnote{We plan to publish all of our source code after the submission under a GPLv3 license.}
To capture all BGP and data-plane probing traffic in the network, we use a custom compiled version of \texttt{tcpdump} using the kernel module \texttt{pf\_ring} in order to handle capturing the large amount of traffic at peak times without any drops.

\subsection{From packet captures to transient violation times}
\label{sec:main:pipeline}
In this section, we outline how to use the information captured for a single sample to both ensure the sample was executed and measured correctly, and to extract transient violation times.

\paragraph{Capture format}
The captured traffic is a regular pcap file produced by \texttt{tcpdump}.
However, when emulating topologies with link delays, it contains all packets sent between routers \emph{twice}, both before and after passing the Delayer.

In addition, we store a so-called \emph{hardware mapping}, allowing to reconstruct which logical router was mapped to which hardware device, on which interfaces it connected to which other router and all the MAC and IP addresses assigned to each network interface.
The source and destination MAC addresses observed on each captured packet allow to uniquely identify which link the packet was currently traversing.
In addition, data-plane probe packets carry a sequence number in their payload.
This allows to uniquely identify each prober packet by its source IP address, destination IP address, and payload sequence number.

\paragraph{Extracting violation times}
Using the unique identification for data-plane probes, we can reconstruct the full forwarding path of any prober packet through the network.
This enables analyzing any path-based forwarding properties, e.g., reachability/isolation of a router, waypoints or stable-path properties.
For analyzing reachability violations, however, it suffices to determine whether a packet reached the last hop or whether it was dropped along the path.

By injecting probe traffic at each router at a constant rate, we get a time series of when packets sent at a source router can reach a destination.
We then approximate the total violation time from such a time series by counting the number of prober packets that got dropped and dividing that number by the probing rate.
Perhaps interestingly, this approach both alleviates the problem of measuring non-consecutive violations, as we can simply accumulate the number of dropped prober packets over time, as well as the synchronization problem if one were to infer transient violation times by correlating the different routers' forwarding states over time, as experienced by real traffic.

\paragraph{Accuracy}
By the Nyquist rate~\cite{oppenheim75digital}, we must sample more often than the ``signal'' changes to avoid information loss.
From our experience and \cite{feldmann04measuring}, the fastest observed reaction times of a single router can be around 2--3\,ms.
Hence, we expect the signal of violation time not to change faster than 1\,ms and, consequently, choose a probing rate of 1000\,pps for our experiments.

\paragraph{Checking a sample for correctness}
To increase the robustness of our measurements, we ensure that we collected the entire convergence process without errors.
For each sample, we first verify that prober packets in the first and last recorded second reach the destination as expected.
This affirms that our testbed always establishes the converged initial state, and waits sufficiently long until the reachability has been restored across all routers.

We further ensure that the Delayers can keep up with all traffic.
Here, we rely on the tofino mirroring each packet twice, once before and once after the Delayer.
This allows us to confirm that the Delayers don't drop any packets, and that they indeed delay the packets by the expected amount.

Finally, for potential packet loss at the collector, we rely on \texttt{tcpdump}'s output of the interface's drop counters, storing them alongside each sample.

\section{Results \& Analysis}
\label{sec:eval}

This section presents a study of violation times from 50 selected network scenarios obtained by our measurement framework after carefully controlling and instrumenting the network.

\subsection{Experimental setup}
\label{sec:eval:setup}

The software components of our measurement framework run on a general-purpose server running Ubuntu 20.04 LTS with 20 physical cores at 2.5\,GHz and a total of 250\,GB of RAM.
At the core of our testbed architecture, we use an Intel Tofino programmable switch featuring 32$\times$\,QSFP28 ports (each supporting 100\,Gbps of traffic), each of which we connect to 4$\times$\,10\,Gbps router interfaces via 1-to-4 DAC breakout cables, or directly to our server (2$\times$\,100\,Gbps QSFP28 DAC, 4$\times$\,10\,Gbps SFP${}^+$\,via a breakout cable).
As routing hardware we employ 12$\times$\,Cisco Nexus9000 C93108TC-FX3P routers, each with 16\,GB of RAM.
In addition, we generate data-plane background traffic (cf.\ \cref{appendix:background-traffic}) on another server running Ubuntu 20.04 LTS with 4 physical cores at 2.4\,GHz and a total of 32\,GB of RAM.
The amount of background traffic is set to 2.5\,Gbps, ensuring a base link utilization of 25\% on each of the routers' 10\,Gbps links.

As described in \cref{sec:main:pipeline}, we measure all experiments at an accuracy of 1\,ms by probing ten randomly chosen prefixes that are affected by the BGP event, thus introducing 10k prober packets (i.e.,\ custom 60\,B IP packets) per second at each router.

\paragraph{Experiment selection}
We expect that transient violation times depend on i)~the information required to share among routers while processing a BGP event (information complexity), ii)~the complexity of processing an event (processing complexity), and iii)~the propagation of the required information (propagation complexity).
The question is \emph{how much} each of these dimensions affects transient violation times.
We take a first step to answering this question for reachability by identifying a set of network parameters that affect each dimension (cf. \cref{table:dimensions_and_params}), and using our framework to control the parameters across experiments.

\setlength{\tabcolsep}{15pt}
\begin{figure}[t]
   \centering
   \begin{tabular}{ll} 
      \toprule
      \textbf{Dimension} & \textbf{Network parameters} \\
      \midrule
      \multirow[t]{2}{*}{Information complexity}   & type of iBGP configuration \\
                                                   & availability of backup routes \\
      \multirow[t]{2}{*}{Processing complexity}    & \# prefixes \\
      \multirow[t]{3}{*}{Propagation complexity}   
                                                   & distance of a router from the event \\
                                                   & distance between the initial and final egress \\
      \bottomrule
    \end{tabular}
    \caption{Evaluated network parameters w.r.t.\ their impact on transient violation times.}
    \label{table:dimensions_and_params}
\end{figure}

\begin{figure*}[t]
  \centering
  \includegraphics{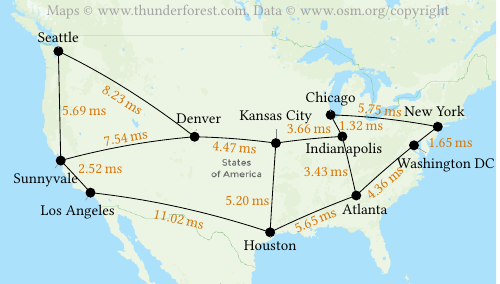}
  \caption{Abilene network topology, including propagation delays between nodes.}
  \label{fig:abilene}
\end{figure*}

\paragraph{Conducting an experiment}
Each experiment is performed as follows:
Initially, the network is in a stable state where all routers can reach all destination prefixes.
At some point, an external BGP event happens.
This causes the network to eventually re-converge to a new state, in which all routers can reach all destination prefixes again.
Unless stated otherwise, each experiment varies one specific network parameter from a default network configuration:
\begin{enumerate}
  \item We use the Abilene network topology (cf.\ \cref{fig:abilene}) from TopologyZoo~\cite{knight11topologyzoo}, i.e., a network spanning the USA with 11~routers.
  \item Los Angeles~(LA) and Kansas City~(KC) are the only border routers connecting directly to external BGP peers.
  \item Link delays are set proportional to the geographic distance of the respective endpoints of a link, i.e., between 1.3\,ms and 11.2\,ms per link (cf.\ \cref{fig:abilene}).
  \item We configure an iBGP full-mesh topology, with equal-IGP-cost links and no internal route-maps.
  \item External BGP events occur at the border router LA, affecting 10k prefixes. KC has a backup route available at all times.
\end{enumerate}

\paragraph{Implementation of BGP events}

We distinguish three BGP event types: announcements, updates changing to a less preferable route than previously, or withraws.
As described in \cref{sec:main:testbed}, we use one of the hardware routers as an external router to inject events into the emulated network at LA.
We trigger the start of an experiment by changing a single outgoing route-map on the external router, depending on the BGP event (permit, AS-path prepend, or deny).

Below we detail the events in the default network configuration.

\subparagraphh{Announce}
Initially, only KC receives a route to the destinations, advertising it to all routers in the network (and LA choosing KC's route as well).
LA advertises a new route with a shorter AS-path, which will eventually be chosen by all routers in the network.
KC withdraws its local route upon learning LA's new route.
All routers eventually choose LA's route.

\subparagraphh{Update-worse}
Initially, both KC and LA receive a route to the destinations, but only LA advertises its route to all routers due to a shorter AS-path (and KC choosing LA's route as well).
LA updates its route with a longer AS-path than KC's route, causing KC to prefer its local route and advertising it to the network.
Upon learning KC's route, LA withdraws its local route.
All routers eventually choose KC's route.

\subparagraphh{Withdraw}
Initially, both KC and LA receive a route to the destinations, but only LA advertises its route to all routers due to a shorter AS-path (and KC choosing LA's route as well).
LA withdraws its route from the network, causing KC to prefer its local route and advertising it to the network.
All routers eventually choose KC's route.

\begin{figure*}[t]
  \centering
  \begin{tikzpicture}

  \newcommand{\MyBoxPlot}[7]{
    \addplot+[
      #1,
      boxplot prepared = {
        lower whisker = #3,
        lower quartile = #4,
        median = #5,
        upper quartile = #6,
        upper whisker = #7,
        draw position = #2,
      },
      ] coordinates {};
  }

  \begin{axis}[
    axis x line = bottom,
    axis y line = left,
    ymin=0,
    ymax=1.9,
    xmin=0.3,
    xmax=3.7,
    width=10cm,
    height=3.8cm,
    xtick={1,2,3},
    xticklabels={
      {announce},
      {update-worse},
      {withdraw},
    },
    ytick={0,0.5,1,1.5,2},
    yticklabels={0\,s,0.5\,s,1\,s,1.5\,s,2,\,s},
    xlabel={},
    ylabel={violation time},
    y label style={anchor=south},
    boxplot/draw direction=y,
    boxplot/box extend=0.35,
    my box/.style={thick, solid},
    legend style={at={(0.5, 1)}, anchor=south, draw=none, /tikz/column 2/.style={column sep=10pt}, inner sep=0pt},
    legend columns=2,
  ]

    \MyBoxPlot{my box,   teal600}{1}{0.0}{0.0}{0.009000000000000001}{0.057999999999999996}{0.1650500000000002} 
    \MyBoxPlot{my box,   rose600}{2}{0.0}{0.0}{0.0}{0.055}{0.183} 
    \MyBoxPlot{my box,   blue600}{3}{0.218}{0.988}{1.307}{1.58}{1.817} 

  \end{axis}
\end{tikzpicture}

  \caption{Withdraw events cause significantly longer violation times compared to announce and update-worse events.}
  \label{fig:events}
\end{figure*}
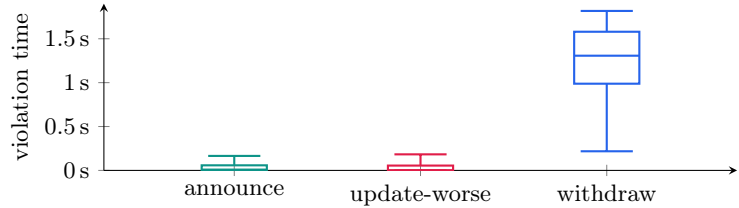

\paragraph{Measuring violation times}
We expect that different BGP events cause reachability violations in different ways.
While withdraw events lead to black holes at the original egress router immediately, other events cause transient forwarding loops or drop packets due to, e.g., reverse-path filtering rules.
Note that forwarding loops can be as short-lived as a few milliseconds difference, which would allow packets to cycle maybe only once and still reach the destination.
Therefore, as proposed in \cref{sec:main:pipeline}, we measure transient reachability violations by relying on the fine-grained data-plane measurements obtained from the prober traffic, only asking whether packets are eventually forwarded to an \emph{active} external router, or not.

Note that, for withdraw events, we consider packets forwarded by the original egress router to its external peer \emph{after} receiving (but before completely processing) a withdraw message as violating packets, i.e., assuming that the external router has already processed the withdraws itself and would thus drop the traffic (is not active anymore).
This defines a clear cut-off when a violation starts, thus rendering experiment results comparable.

\paragraph{Sampling}
In the following, each boxplot refers to a given network scenario measured for one BGP event and summarizes violation times from at least 50 samples, yielding a total of 5.5k data points for measurements across all routers (11$\times$\,ten randomly probed prefixes per sample/experiment run).
In our paper, boxplots show the 5th, 25th, 50th, 75th, and 95th percentile of violation times.

\paragraph{Baseline}
\Cref{fig:events} shows the violation times for different BGP events in the default network setting.
As expected, we observe that route withdrawals consistently cause the longest reachability violation times, significantly surpassing the other event types.
Given this observation, we focus our subsequent analysis primarily on route withdrawal events, while mentioning selected interesting effects observed for update-worse events in \cref{sec:eval:information}.

\subsection{Information complexity}
\label{sec:eval:information}

The first dimension we explore is the amount of information that must be exchanged to converge to the new stable state.
For instance, all routers in the initial state may already know the routes that will be selected in the final state.
Then, routers must only propagate the information of the event through the network.
However, these routes may be \textit{hidden}, i.e., they must first be distributed through the network before they can be selected.
This information may not be propagated directly, but via a route reflector which adds an indirection layer.
While route reflection allows to decrease the required routing memory, it can also increase the convergence time~\cite{park11quantifying} and violation time, as we show.
Interestingly, for update-worse events the indirection introduced by route reflectors can also help to \textbf{decrease the violation times} experienced in the network!

\paragraph{Withdraw events}
When routes are withdrawn, the border router initially receiving those events will immediately start dropping packets.
The border router then propagates the information, and consequently the black hole, in the network.
These black holes are then resolved once the backup route is propagated.
In the following, we analyze the effect of the iBGP topology and the visibility of the backup route on the violation time.

\begin{figure*}[t]
  \centering
  \begin{tikzpicture}

  \newcommand{\MyBoxPlot}[7]{
    \addplot+[
      #1,
      boxplot prepared = {
        lower whisker = #3,
        lower quartile = #4,
        median = #5,
        upper quartile = #6,
        upper whisker = #7,
        draw position = #2,
      },
      ] coordinates {};
  }

  \begin{axis}[
    axis x line = center,
    axis y line = center,
    ymin=0,
    ymax=5,
    xmin=0,
    xmax=7.4,
    width=\linewidth,
    height=5cm,
    xtick={1,2,2.6,3.2,4.2,4.8,5.4,6.4},
    xticklabels={full mesh, Atlanta, New York, Seattle, Atlanta+NewYork,Atlanta+Seattle,NewYork+Seattle,Atlanta+NewYork+Seattle},
    ytick={0,1,2,3,4},
    yticklabels={0\,s,1\,s,2\,s,3\,s,4\,s},
    xlabel={route reflectors},
    ylabel={violation time},
    y label style={anchor=south},
    x tick label style={rotate=25,anchor=east, outer sep=5pt, xshift=2mm, yshift=-1mm},
    boxplot/draw direction=y,
    boxplot/box extend=0.28,
    my box/.style={thick, solid},
    legend style={at={(0.5, 1)}, anchor=south, draw=none, /tikz/column 2/.style={column sep=10pt}, inner sep=0pt},
    legend columns=2,
  ]

    \MyBoxPlot{my box,   teal600}{1}{0.218}{0.988}{1.307}{1.58}{1.817}
    \MyBoxPlot{my box,   blue600}{2}{0.6565500000000001}{1.8095}{2.273}{2.68125}{2.958}
    \MyBoxPlot{my box,   blue600}{2.6}{0.407}{1.80475}{2.3605}{2.78675}{3.09755}
    \MyBoxPlot{my box,   blue600}{3.2}{0.5786}{1.78375}{2.1745}{2.65375}{3.038}
    \MyBoxPlot{my box, purple600}{4.2}{0.78805}{1.92}{2.7445000000000004}{3.513}{3.79275}
    \MyBoxPlot{my box, purple600}{4.8}{1.16845}{2.35125}{2.9625000000000004}{3.33275}{3.7164999999999986}
    \MyBoxPlot{my box, purple600}{5.4}{0.753}{1.86725}{2.507}{3.026}{3.4705500000000002}
    \MyBoxPlot{my box,   rose600}{6.4}{1.39815}{2.16575}{2.8395}{3.49575}{4.185}

  \end{axis}
\end{tikzpicture}

  \caption{Route reflection increases the violation time for withdraw events.}
  \label{fig:effect_of_rr}
\end{figure*}
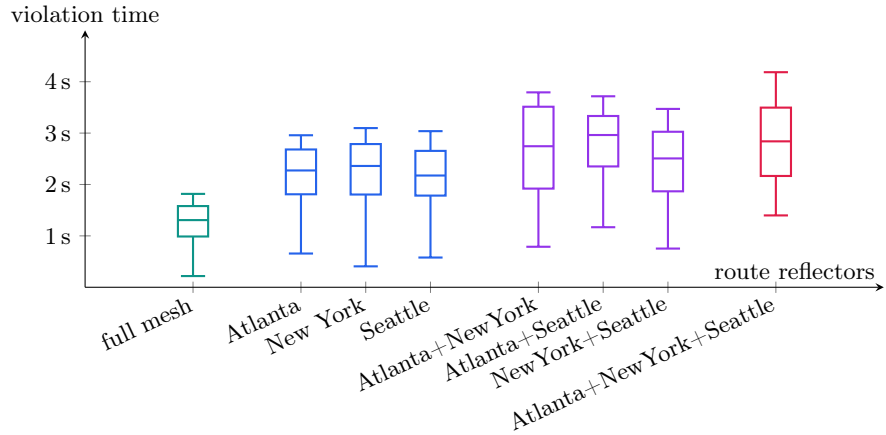
\subparagraphh{Full-mesh vs.\ route reflection}
As our first experiment, we compare the effects on violation times between a full-mesh and several route-reflection configurations.
We configure between one and three route reflectors, at different locations in the network, and set up all other routers as route-reflector clients to all route reflectors.
This includes the route reflectors themselves, thus also communicating their currently selected best routes among each other.

We use our default network parameters and showcase the results for a BGP withdraw event, observing the violation times shown in \cref{fig:effect_of_rr}.
Each boxplot corresponds to a full-mesh or route-reflection configuration.
We observe that the median violation time increases from 1.3\,s for full-mesh to on average 2.3\,s/2.7\,s/2.8\,s for one/two/three reflectors.
The increase is expected, due to the indirection introduced by route reflectors.
Further, we see that using route reflectors approximately doubles the violation time compared to the full-mesh scenario:
Route reflectors initially react to a withdrawal by sending another update message, advertising the route received from another reflector which has not been withdrawn yet.
Only after observing the withdrawal from the other route reflector(s), they propagate the event information to all other routers in the network, leading to a backup route being advertised and eventually propagated to all routers to resolve the violation.
This behavior amplifies slightly as we configure more route reflectors.

Overall, the use of route reflectors results in approximately double the violation time compared to the full-mesh scenario.
We attribute this to the increased number of BGP processing steps on the critical path towards resolving the violation, as the route reflectors need to replicate both actions performed by the originator of the event and the backup router in this scenario.

\subparagraphh{Hidden vs.\ visible backup routes}
Next, we study the effects of backup routes being available at all routers in the network.
Intuitively, considering withdraw events, it may seem obvious that violation times decrease when a backup route is continuously advertised to all routers~\cite{shrieck2010bgp}.
This is because it then suffices to propagate the withdraw event through the network and routers can locally resolve the violation by choosing the available backup route.
Nonetheless, we verify empirically that this is, indeed, the case.
In addition, we configure up to two route reflectors, and distinguish whether the backup route is hidden to the route reflectors, available at both route reflectors, or even chosen by one of the route reflectors (and thus advertised to all routers).
Similar to above, we set up all other routers as route-reflector clients to all route reflectors, including the route reflectors themselves.

We use our default network parameters and configure a backup route in KC with the same AS-path length as the initial route in LA.
Introducing BGP withdraw events at LA, we expect violation times to decrease by keeping a backup route advertised, as mentioned above.
In combination with route reflection, a route reflector essentially functions as the backup router in the full-mesh case, i.e., advertising the backup route to all routers upon learning the withdraw.
Hence, we expect similar violations here as in the full-mesh case with a hidden backup, although slightly faster because the originator of the event, LA, only has to communicate the withdraws to the route reflector instead of all routers.

As some routers already prefer the available backup route initially, they do not experience a violation at all.
Thus, we subsequently discuss the violations experienced by the router Houston for ten randomly selected probed prefixes, exemplary for any router that switches its egress due to the event.\footnote{We choose to present a single router's measurements as it is easier to interpret the behavior for one concrete example. Other routers exhibit similar behavior.}

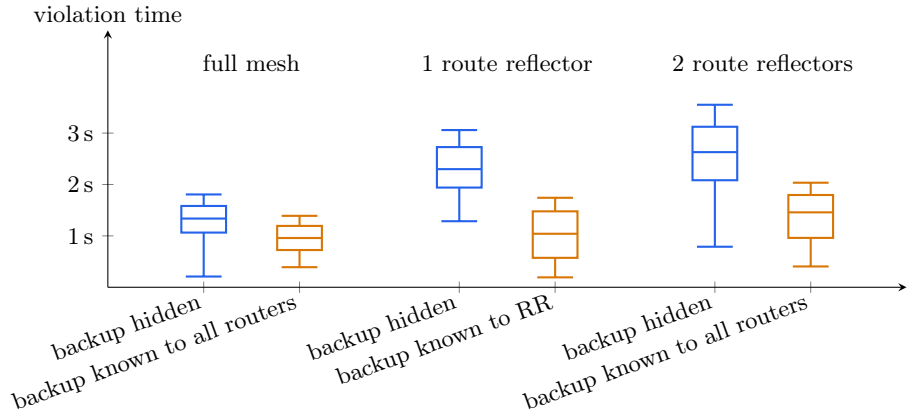
\begin{figure*}[t]
  \centering
  \begin{tikzpicture}

  \newcommand{\MyBoxPlot}[7]{
    \addplot+[
      #1,
      boxplot prepared = {
        lower whisker = #3,
        lower quartile = #4,
        median = #5,
        upper quartile = #6,
        upper whisker = #7,
        draw position = #2,
      },
      ] coordinates {};
  }

  \begin{axis}[
    axis x line = center,
    axis y line = center,
    ymin=0,
    ymax=5,
    xmin=0.4,
    xmax=5.4,
    width=\linewidth,
    height=5cm,
    xtick={1,1.6,2.6,3.2,4.2,4.8},
    xticklabels={
      {backup hidden},
      {backup known to all routers},
      {backup hidden},
      {backup known to RR},
      {backup hidden},
      {backup known to all routers},
    },
    ytick={0,1,2,3},
    yticklabels={0\,s,1\,s,2\,s,3\,s},
    xlabel={},
    ylabel={violation time},
    y label style={anchor=south},
    x tick label style={rotate=20,anchor=east, outer sep=5pt, xshift=2mm, yshift=-1mm},
    boxplot/draw direction=y,
    boxplot/box extend=0.28,
    my box/.style={thick, solid},
    legend style={at={(0.5, 1)}, anchor=south, draw=none, /tikz/column 2/.style={column sep=10pt}, inner sep=0pt},
    legend columns=2,
  ]

    \MyBoxPlot{my box,  blue600}{1}{0.209}{1.064}{1.337}{1.582}{1.8061} 
    \MyBoxPlot{my box, amber600}{1.6}{0.39070000000000005}{0.7244999999999999}{0.9584999999999999}{1.19375}{1.3899499999999996} 
    \MyBoxPlot{my box,  blue600}{2.6}{1.28535}{1.94}{2.2984999999999998}{2.72775}{3.0591} 
    \MyBoxPlot{my box, amber600}{3.2}{0.19285000000000002}{0.5725}{1.041}{1.4775}{1.7412} 
    \MyBoxPlot{my box,  blue600}{4.2}{0.7886000000000001}{2.08275}{2.6295}{3.12325}{3.5513999999999997} 
    \MyBoxPlot{my box, amber600}{4.8}{0.40475000000000005}{0.9604999999999999}{1.4580000000000002}{1.7955}{2.0345} 

    \node[inner sep=4pt, above] at (axis cs:1.3, 4) {full mesh};
    \node[inner sep=4pt, above] at (axis cs:2.9, 4) {1 route reflector};
    \node[inner sep=4pt, above] at (axis cs:4.5, 4) {2 route reflectors};

  \end{axis}
\end{tikzpicture}

  \caption{For withdraw events, continuously advertising a backup route significantly reduces the violation times.}
  \label{fig:effect_of_hidden_backup}
\end{figure*}

We observe the violation times shown in \cref{fig:effect_of_hidden_backup}.
We verify that all of our expectations are met.
To begin with, we observe that the violation time in the full-mesh case roughly halves as we keep the backup route continuously advertised.
We attribute this to the reduced number of BGP processing steps on the critical path towards resolving the violation.
Further, we continue to observe halved violation times when adding route reflectors.

\paragraph{Update-worse events}
While for withdraw events, the transient violations come generally from transient black holes at some routers, those are far less common for other events.
Instead, the \emph{timing} of BGP messages sent and processed on different routers is far more important: the order in which two neighboring routers change their next hop determines whether forwarding loops or packet drops due to reverse path filtering occur.

Both announce and update-worse events are triggered by BGP update messages.
The main difference is that the update-worse evnet requires path exploration to find the backup route.
In the following, we analyze the effect of route-reflection on the violation time upon announce and update-worse events, and on the effect of backup-route visibility upon update-worse events only.

\subparagraphh{Full-mesh vs.\ route reflection}
When the AS-path for an existing route increases in length, the border router first disseminates the update to other routers without changing its local forwarding decisions.
Once this information reaches the backup router it will start selecting and propagating its local backup route, allowing other routers to change as well.
In a full-mesh, these updates traverse in the opposite direction of the forwarding, and thus, minimize the change of forwarding loops to occur.
However, route reflection breaks this property; the information can reach the routers closest to the new destination last.
We thus expect forwarding loops to appear more frequent and for a longer time with route reflection than in an iBGP full-mesh.

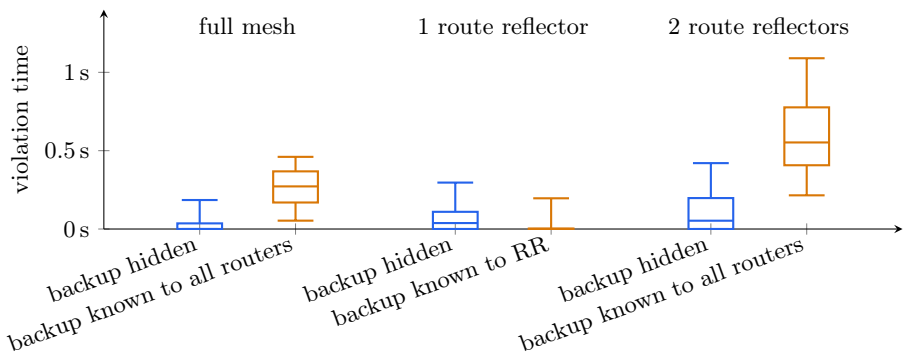
\begin{figure*}[t]
  \centering
  \begin{tikzpicture}

  \newcommand{\MyBoxPlot}[7]{
    \addplot+[
      #1,
      boxplot prepared = {
        lower whisker = #3,
        lower quartile = #4,
        median = #5,
        upper quartile = #6,
        upper whisker = #7,
        draw position = #2,
      },
      ] coordinates {};
  }

  \begin{axis}[
    axis x line = bottom,
    axis y line = left,
    ymin=0,
    ymax=1.4,
    xmin=0.4,
    xmax=5.4,
    width=\linewidth,
    height=4.5cm,
    xtick={1,1.6,2.6,3.2,4.2,4.8},
    xticklabels={
      {backup hidden},
      {backup known to all routers},
      {backup hidden},
      {backup known to RR},
      {backup hidden},
      {backup known to all routers},
    },
    ytick={0,0.5,1,1.5},
    yticklabels={0\,s,0.5\,s,1\,s,1.5\,s},
    xlabel={},
    ylabel={violation time},
    y label style={anchor=south},
    x tick label style={rotate=20,anchor=east, outer sep=5pt, xshift=2mm, yshift=-1mm},
    boxplot/draw direction=y,
    boxplot/box extend=0.28,
    my box/.style={thick, solid},
    legend style={at={(0.5, 1)}, anchor=south, draw=none, /tikz/column 2/.style={column sep=10pt}, inner sep=0pt},
    legend columns=2,
  ]

    \MyBoxPlot{my box,  blue600}{1}{0.0}{0.0}{0.0}{0.036}{0.18504999999999996} 
    \MyBoxPlot{my box, amber600}{1.6}{0.05335000000000001}{0.16925}{0.272}{0.368}{0.46054999999999996} 
    \MyBoxPlot{my box,  blue600}{2.6}{0.0}{0.0}{0.038}{0.10975}{0.2963000000000003} 
    \MyBoxPlot{my box, amber600}{3.2}{0.0}{0.0}{0.0}{0.003}{0.19610000000000002} 
    \MyBoxPlot{my box,  blue600}{4.2}{0.0}{0.0}{0.053}{0.1975}{0.42055} 
    \MyBoxPlot{my box, amber600}{4.8}{0.215}{0.407}{0.5525}{0.7765}{1.09} 

    \node at (axis cs:1.3, 1.3) {full mesh};
    \node at (axis cs:2.9, 1.3) {1 route reflector};
    \node at (axis cs:4.5, 1.3) {2 route reflectors};

  \end{axis}
\end{tikzpicture}

  \caption{For updates-worse events, continuously advertising a backup route can significantly \textbf{increase} the observed violation times.}
  \label{fig:effect_of_hidden_backup_updateworse}
\end{figure*}

The results are shown in \cref{fig:effect_of_hidden_backup_updateworse}, where the blue bars show the violation times for full-mesh and for one and two route reflectors.
We confirm that the trend we described goes up as expected, measuring median violation times of 0\,ms, 38\,ms and 53\,ms.
Further, we confirmed manually for a few samples that, indeed, drops happen due to forwarding loops and mainly reverse-path filtering.

\subparagraphh{Hidden vs.\ visible backup routes}
To study the effect of hidden and visible backup routes in the update-worse scenario, we repeat the experiment as for the withdrawal event, where KC learns initially learns backup routes with either a longer (hidden) or the same (visible) AS-path length as LA.
We trigger events that increase the AS-path length of routes announced to LA, causing the network to use the backup routes at KC.
We then measure how the visibility of KC's routes affect the violation time.

In the full-mesh scenario with a hidden backup route, we expect that the transient violations are significantly smaller than in the withdraw case, as LA continues to forward incoming traffic toward the destination successfully until it eventually learns the backup route from KC like all other routers (cf.\ \cref{fig:events}).
With the backup route continuously advertised, however, it becomes more interesting.
As LA sees KC's route throughout the experiment, we expect LA to withdraw its route and directly choose KC as its next hop.
While not actively dropping traffic due to a transient black hole in the routing state, LA should now drop all incoming traffic for the affected destination prefixes from its neighbor Houston due to reverse path filtering.
Consequently and perhaps counterintuitively, we expect the violation times to \emph{increase} when backup routes are available at all routers in the network.

Following this argument, we expect to see a \emph{decrease} in violation times as we introduce route reflectors in the network.
That is, because route reflectors (typically) only select and advertise a single best route, thus hiding the backup route from LA initially and avoiding the withdraw to occur.

Consider the violation times experienced at the router Houston (cf.\ \cref{fig:effect_of_hidden_backup_updateworse}).
For the iBGP full-mesh case, the violation time \emph{increases} significantly when the backup routes from KC are known to all routers, namely from a median of 0\,ms violation to 169\,ms violation.
Note that this number does not match the violation for a withdraw event, as any traffic forwarded to LA's external router during LA's initial processing is now attributed as non-violating traffic, as the external router knows valid (although less preferred) routes to the destinations.

Next, we compare the violation times if the backup route at KC is hidden/visible in the scenario with one route reflector at Seattle (SE).
In both cases, SE initially prefers LA's route.
As discussed above, we expect to see a higher violation time than in the full-mesh case when the backup route is hidden.
When the backup route is available (to the route reflector), however, SE can directly distribute the backup route.
In other words, the route reflector essentially functions as the backup router in the full-mesh case, explaining why the median violation observed goes back from 38\,ms down to 0\,ms.

Finally, we examine the network with two route reflectors, one at SE and one at New York (NY).
The scenario is as for one route reflector, but now SE initially prefers the route from LA, while NY prefers the route from KC due to its lower IGP cost on that path.
Hence, NY continuously advertises KC's backup route, making both routes available to all routers.
Similarly to the case with a visible backup route in the iBGP full-mesh case, the violation time significantly increases from 53\,ms to 552\,ms when the backup route is already visible to LA.

Notably, we observe even higher violation times than in this full-mesh case, because of the indirection imposed by the route reflectors.
More precisely, we attribute the violations not only to an extra hop in the propagation of information, but because the route reflector SE, upon learning LA's withdraw, sends BGP updates advertising KC's backup route to all other routers in the network.
This, however, is much slower than simply relaying withdraw messages and does not allow routers to make a similarly quick transition to KC's backup route.

\subsection{Processing complexity}
\label{sec:eval:processing}

Next, we want to measure the router's processing time overhead.
While the processing intensity of a single prefix can also vary depending on the complexity of installed route maps, available number of peers, or the number of alternate routes to compare against, we find that the only factor that can scale almost limitless is the sheer number of prefixes affected by an event.
In addition, this yields the unique benefit of observing the progress of the processing, as each updated prefix can be observed on the network already.
Thus, we explore the impact of the processing complexity on transient violation times by varying the number of prefixes affected by an event.

\paragraph{Number of prefixes}
We use the default parameters except for the number of prefixes involved in the BGP events, which we vary from one to 1 million prefixes.
More precisely, we experiment with the following steps: 1, 2, 5, 10, 25, 50, 100, 250, 500, 1k, 2.5k, 5k, 10k, 25k, 50k, 100k, 250k, 500k, and 1M.

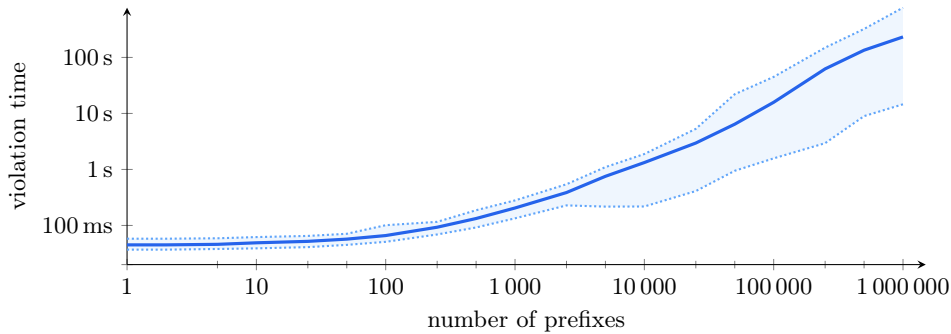
\begin{figure*}[t]
  \centering
  \begin{tikzpicture}

\newcommand{\WithdrawFileName}{data/prefixes_Abilene_ExtLosAngelesKansasCity_FullMesh_Prefix1__PhysicalExternalWithdrawAllPrefixesAtLosAngeles.csv}

  \begin{axis}[
    axis x line = bottom,
    axis y line = left,
    xmode=log,
    ymode=log,
    width=\linewidth,
    height=5cm,
    xlabel={number of prefixes},
    ylabel={violation time},
    y label style={anchor=south},
    xmin=1,
    xmax=1500000,
    xtick={1,10,100,1000,10000,100000,1000000},
    xticklabels={1,10,100,1\,000,10\,000,100\,000,1\,000\,000},
    minor xtick={5,25,50,250,500,2500,5000,25000,50000,250000,500000},
    ymin=0.02,
    ytick={0.02, 0.1, 1, 10, 100},
    yticklabels={, 100\,ms, 1\,s, 10\,s, 100\,s},
  ]

    \addplot+[mark=none, thick, draw=blue400, densely dotted, name path=upper] table[x=num_prefixes, y=q95, col sep=comma] {\WithdrawFileName};
    \addplot+[mark=none, thick, draw=blue400, densely dotted, name path=lower] table[x=num_prefixes, y=q5, col sep=comma] {\WithdrawFileName};
    \addplot [draw=none, fill=blue400, fill opacity=0.1] fill between[of=lower and upper];
    \addplot+[very thick, mark=none, blue600] table[x=num_prefixes, y=q50, col sep=comma] {\WithdrawFileName};

  \end{axis}
\end{tikzpicture}

  \caption{Violation times are dominated by the BGP processing for large BGP events. \textit{(Note that the 5\%/95\%-percentiles appear skewed due to the log-scale on the y-axis.)}}
  \label{fig:effect_of_prefixes}
\end{figure*}

We expect that a withdraw event affecting a large number of prefixes all at once leads to longer violation times, as each prefix' update has to be computed individually, and BGP messages may thus queue up at the routers while awaiting processing.
This generally slows down the routers' reactions to messages, thus prolonging transient black holes occurring.
To that end, recall that we consider packets forwarded by the original egress router to its external peer \emph{after} receiving (but before completely processing) a withdraw message as violating packets.

\Cref{fig:effect_of_prefixes} shows a log-scale graph depicting the median violation times, with areas above and below showing the 95th and 5th percentiles, respectively.
Note that the upper percentiles appear smaller than the lower ones due to the log-scale on the y-axis.
For Withdraw events, we observe a clear increase in violation times as we increase the number of prefixes:
The median violation time varies from only 45\,ms for a single prefix to 216\,s for 1M prefixes.
This dramatic increase highlights the significant role that control-plane processing complexity plays in determining violation times.
For sufficiently large numbers of prefixes, this dimension dominates violation times as they increase linearly with the number of prefixes.
Further, we see that more prefixes lead to more variable violation times, because the range of processing/queueing times increases.

\subsection{Propagation complexity}\label{sec:eval:propagation}

Last, we consider the role of propagation delays in transient violation times.
While the propagation required by an event may clearly affect the transient violation time, we expect this to cause effects in the order of (tens of) milliseconds.
Hence, one may neglect such effects for large events, while proportionally more relevant for smaller events.
In this section, we briefly verify our intuition about violation times depending on the propagation delays using events affecting only 100 prefixes.

\paragraph{Total propagation delay}
In a full-mesh scenario, the propagation of an event has three parts: i)~the propagation of any in-flight traffic (cf.\ \cref{sec:motivation}) on the path from the router whose violation we consider to the original egress, ii)~the propagation from the original egress to the backup router, and iii)~the propagation from the backup router to advertise its new route to our considered router at hand.
In the following, we will refer to the sum of these three values as the \emph{total propagation delay} for a router in a given scenario, and show how it correlates with the experienced transient violations.

\paragraph{Router's distance from the event}
We expect a router's location in the network to play a role in determining its violation.
The first router to recover will be the egress of the backup route itself, i.e., KC.
Its violation is determined by doubling its propagation delay from the original egress (due to traffic in-flight being dropped and traffic sent until the event has been propagated to the backup router), plus two processing delays, one at LA and another at KC.

All other routers, however, will experience longer violation times than the backup router as they have to additionally wait for KC to disseminate the backup route.
For any other router~$r$, we expect the violation time to be roughly equal to its total propagation delay, i.e., from $r$ to LA (accounting for in-flight packets being dropped), from LA to KC, and from KC to $r$, plus three additional processing delays.

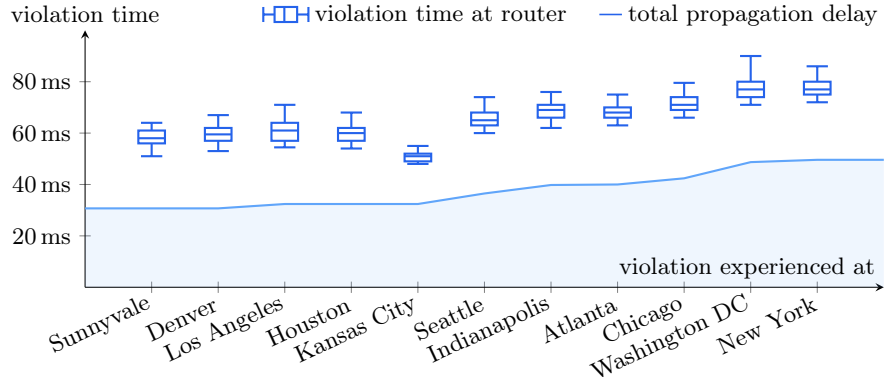
\begin{figure*}[t]
  \centering
  \begin{tikzpicture}[
    combined distance/.style={anchor=north, font=\itshape},
  ]

  \newcommand{\MyBoxPlot}[7]{
    \addplot+[
      #1,
      boxplot prepared = {
        lower whisker = #3,
        lower quartile = #4,
        median = #5,
        upper quartile = #6,
        upper whisker = #7,
        draw position = #2,
      },
      ] coordinates {};
  }

  \pgfplotsset{
    short legend/.style={%
      legend image code/.code={
        \draw[##1,line width=0.6pt]
          plot coordinates {
            (0cm,0cm)
            (0.3cm,0cm)
          };%
      }
    }
  }

  \begin{axis}[
    axis x line = center,
    axis y line = center,
    ymin=0,
    ymax=0.1,
    xmin=0,
    xmax=12,
    width=\linewidth,
    height=5cm,
    xtick={1,2,3,4,5,6,7,8,9,10,11},
    xticklabels={
      {Sunnyvale},
      {Denver},
      {Los Angeles},
      {Houston},
      {Kansas City},
      {Seattle},
      {Indianapolis},
      {Atlanta},
      {Chicago},
      {Washington DC},
      {New York},
    },
    ytick={0.02,0.04,0.06,0.08},
    yticklabels={20\,ms,40\,ms,60\,ms,80\,ms},
    xlabel={violation experienced at},
    ylabel={violation time},
    y label style={anchor=south},
    x tick label style={rotate=25,anchor=east, outer sep=5pt, xshift=2mm, yshift=-1mm},
    boxplot/draw direction=y,
    boxplot/box extend=0.4,
    router/.style={thick, solid, blue600},
    legend style={at={(1, 1)}, anchor=south east, draw=none, /tikz/column 2/.style={column sep=10pt}, inner sep=0pt},
    legend columns=2,
  ]
    \addlegendimage{box plot legend={thick, blue600}}
    \addlegendentry{violation time at router}
    \addlegendimage{short legend,thick,blue600}
    \addlegendentry{total propagation delay}

    \MyBoxPlot{router}{1}{0.051}{0.055999999999999994}{0.057999999999999996}{0.061}{0.064}

    \MyBoxPlot{router}{2}{0.053}{0.057}{0.0595}{0.062}{0.067}

    \MyBoxPlot{router}{3}{0.05445000000000001}{0.057}{0.061}{0.064}{0.071}

    \MyBoxPlot{router}{4}{0.054000000000000006}{0.057}{0.06}{0.062}{0.068}

    \MyBoxPlot{router}{5}{0.048}{0.049}{0.051}{0.052000000000000005}{0.055}

    \MyBoxPlot{router}{6}{0.06}{0.063}{0.065}{0.068}{0.07400000000000001}

    \MyBoxPlot{router}{7}{0.062}{0.066}{0.069}{0.071}{0.076}

    \MyBoxPlot{router}{8}{0.063}{0.066}{0.068}{0.07}{0.075}

    \MyBoxPlot{router}{9}{0.066}{0.069}{0.071}{0.07400000000000001}{0.07954999999999995}

    \MyBoxPlot{router}{10}{0.071}{0.07400000000000001}{0.077}{0.08}{0.09}

    \MyBoxPlot{router}{11}{0.07200000000000001}{0.075}{0.077}{0.08}{0.086}

    \addplot+[mark=none, draw=none, densely dotted, name path=zero] coordinates {
      (0,0)
      (12,0)
    };
    \addplot+[mark=none, thick, draw=blue400, name path=distances] coordinates {
        (0,0.0307)
        (1,0.0307)
        (2,0.0307)
        (3,0.0324)
        (4,0.0324)
        (5,0.0324)
        (6,0.0365)
        (7,0.0398)
        (8,0.0400)
        (9,0.0424)
        (10,0.0487)
        (11,0.0496)
        (12,0.0496)
    };
    \addplot [draw=none, fill=blue400, fill opacity=0.1] fill between[of=zero and distances];

  \end{axis}
\end{tikzpicture}

  \caption{The violation time for a withdraw event measured at all routers in the network. The line represents the propagation delay from that router to LA, then KC, and back.}
  \label{fig:effect_of_router_distance}
\end{figure*}

\cref{fig:effect_of_router_distance} shows the measured violation times at all routers in the network.
Indeed, it confirms that KC has the smallest violation time.
Further, it confirms that the violation time of all other routers follows the order of their total propagation delay with respect to LA and KC, as described above.
Considering the violation time at KC and its total propagation delay of 32.44\,ms and a median violation of 51\,ms, we can infer the processing time for 100 prefixes to be around 9--10\,ms.
This then allows to estimate the violation time, e.g., at Sunnyvale: its median violation time of 58\,ms is roughly equal to its total propagation delay of 30.8\,ms, plus three processing delays.
Similarly, for all cases confirm that the median of the observed transient violation lies within three processing delays between 9.1\,ms and 9.8\,ms of the total propagation delay, i.e., accurately predicting the median violation times with an error of $\pm$ 1.2\,ms.

\paragraph{Distance between the initial and final egress}

For a given router $r$, increasing only the distance between the initial and backup router by $x$ will further delay the notification of the backup router by $x$, hence increase $r$'s violation time by $x$.
In a given topology though, moving the backup router also affects another component of the total propagation delay: the propagation delay between $r$ and the backup may increase or decrease by $y$.
Hence, we expect the overall change in the violation time to be $x\pm y$.

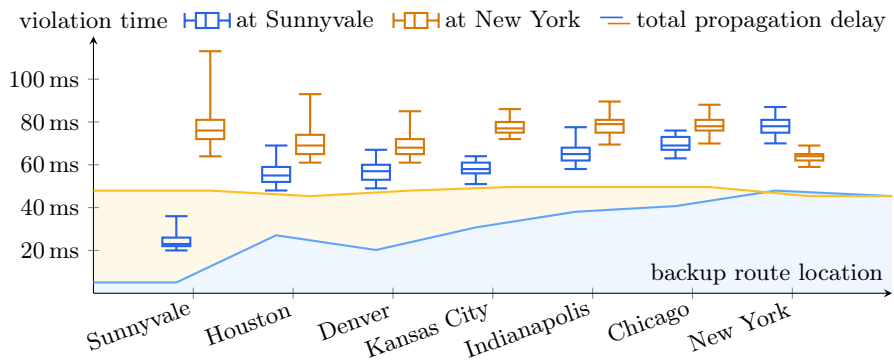
\begin{figure*}[t]
  \centering
  \begin{tikzpicture}

  \newcommand{\MyBoxPlot}[7]{
    \addplot+[
      #1,
      boxplot prepared = {
        lower whisker = #3,
        lower quartile = #4,
        median = #5,
        upper quartile = #6,
        upper whisker = #7,
        draw position = #2,
      },
      ] coordinates {};
  }

  \pgfplotsset{
    custom legend/.style={%
      legend image code/.code={
        \draw[blue600,line width=0.6pt]
          plot coordinates {
            (0cm,.3ex)
            (0.3cm,.3ex)
          };%
        \draw[amber600,line width=0.6pt]
          plot coordinates {
            (0.1cm,-.3ex)
            (0.4cm,-.3ex)
          };%
      }
    }
  }

  \begin{axis}[
    axis x line = center,
    axis y line = center,
    ymin=0,
    ymax=0.12,
    xmin=0,
    xmax=8,
    width=\linewidth,
    height=5cm,
    xtick={1,2,3,4,5,6,7},
    xticklabels={Sunnyvale, Houston, Denver, Kansas City, Indianapolis, Chicago, New York},
    ytick={0, 0.02, 0.04, 0.06, 0.08,0.1},
    yticklabels={0,20\,ms,40\,ms,60\,ms,80\,ms,100\,ms},
    xlabel={backup route location},
    ylabel={violation time},
    y label style={anchor=south},
    x tick label style={rotate=20,anchor=east, outer sep=5pt, xshift=2mm, yshift=-1mm},
    boxplot/draw direction=y,
    boxplot/box extend=0.28,
    sunnyvale/.style={thick, solid, blue600},
    new york/.style={thick, solid, amber600},
    legend style={at={(1, 1)}, anchor=south east, draw=none, /tikz/column 2/.style={column sep=5pt}, /tikz/column 4/.style={column sep=5pt}, inner sep=0pt},
    legend columns=-1,
  ]

    \addlegendimage{box plot legend={thick, blue600}}
    \addlegendentry{at Sunnyvale}
    \addlegendimage{box plot legend={thick, amber600}}
    \addlegendentry{at New York}
    \addlegendimage{custom legend}
    \addlegendentry{total propagation delay}

    \MyBoxPlot{sunnyvale}{0.83}{0.02}{0.022000000000000002}{0.023}{0.026000000000000002}{0.036000000000000004}
    \MyBoxPlot{new york }{1.17}{0.06395}{0.07200000000000001}{0.076}{0.081}{0.11304999999999996}

    \MyBoxPlot{sunnyvale}{1.83}{0.048}{0.052000000000000005}{0.055}{0.059000000000000004}{0.069}
    \MyBoxPlot{new york }{2.17}{0.061}{0.065}{0.069}{0.07400000000000001}{0.09300000000000001}

    \MyBoxPlot{sunnyvale}{2.83}{0.049}{0.053}{0.057}{0.06}{0.067}
    \MyBoxPlot{new york }{3.17}{0.061}{0.065}{0.068}{0.07200000000000001}{0.085}

    \MyBoxPlot{sunnyvale}{3.83}{0.051}{0.055999999999999994}{0.057999999999999996}{0.061}{0.064}
    \MyBoxPlot{new york }{4.17}{0.07200000000000001}{0.075}{0.077}{0.08}{0.086}

    \MyBoxPlot{sunnyvale}{4.83}{0.057999999999999996}{0.062}{0.065}{0.068}{0.07754999999999995}
    \MyBoxPlot{new york }{5.17}{0.06945000000000001}{0.075}{0.079}{0.081}{0.08954999999999996}

    \MyBoxPlot{sunnyvale}{5.83}{0.063}{0.067}{0.069}{0.073}{0.076}
    \MyBoxPlot{new york }{6.17}{0.06995000000000001}{0.076}{0.078}{0.081}{0.08804999999999996}

    \MyBoxPlot{sunnyvale}{6.83}{0.07}{0.075}{0.078}{0.081}{0.087}
    \MyBoxPlot{new york }{7.17}{0.059000000000000004}{0.062}{0.064}{0.065}{0.069}

    \addplot[mark=none, draw=none, densely dotted, name path=zero] coordinates {
      (0,0)
      (8,0)
    };
    \addplot[mark=none, thick, draw=blue400, name path=distances from sv] coordinates {
        (0,0.00504)
        (0.83,0.00504) 
        (1.83,0.02708) 
        (2.83,0.02021) 
        (3.83,0.03075) 
        (4.83,0.03807) 
        (5.83,0.04071) 
        (6.83,0.04794) 
        (8,0.04536)
    };
    \addplot[draw=none, fill=blue400, fill opacity=0.1] fill between[of=zero and distances from sv];
    \addplot[mark=none, thick, draw=amber400, name path=distances from ny] coordinates {
        (0,0.04794)
        (1.17,0.04794) 
        (2.17,0.04536) 
        (3.17,0.04794) 
        (4.17,0.04963) 
        (5.17,0.04963) 
        (6.17,0.04963) 
        (7.17,0.04536) 
        (8,0.04536)
    };
    \addplot[draw=none, fill=amber400, fill opacity=0.1] fill between[of=distances from sv and distances from ny];

  \end{axis}
\end{tikzpicture}

  \caption{Violatoin time measured at Sunnyvale (blue) and New York (orange), where the backup route is available at different locations in the network.}
  \label{fig:effect_of_distance}
\end{figure*}

\cref{fig:effect_of_distance} shows measured violations for Sunnyvale (SV) and New York (NY) when the initial egress is at LA.
We demonstrate the effect of different backup routers on the x-axis.
Consider the blue boxplots, which show SV's violation times.
We observe that moving the backup router from HS to NY increases the median violation time from 52\,ms to 75\,ms, because the total propagation time also increases by approximately the same amount, from 27\,ms to 48\,ms.
Note that when the backup is at SV itself, SV's median violation time of 22\,ms is almost halved compared to next longest violation time at HS.
The reason is that when SV is both the router for which we measure violations and the backup one, switching to the new route saves one processing time plus the time for the backup router to disemminate the route.
In addition, its mere proximity to the original egress LA accounts for a total propagation delay of only 5\,ms.

The orange boxplots show similar results for NY's observed violation times.
The main difference is that moving the backup route (from SV to CH) not always increases the median violation time.
Instead, the violation time increases or decreases by as much as total propagation delay increases or decreases.

\section{Related work}

This section presents an overview of other works that apply to the general area of network convergence, and highlights the key differences from our approach.
We discuss these works in decreasing order of similarity to our presented approach.

\paragraph{BGP transient violations on specific setups}
The closest work also analyzes transient violations due to external routing events, but is limited in terms of routing events and network configurations.
Bush et al.~\cite{bush05happy} withdraw (announce) a multi-homed host's prefix from (to) one of two providers and measure the degradation in end-to-end performance of traffic from geodistributed probes to the host.
They find that the degradation is not significant and cannot be captured by control-plane signals (e.g., from RouteViews).
Wang et al.~\cite{wang06ameasurementstudy} follow a similar methodology and additionally correlate packet loss with traceroute and ping responses to attribute loss to routing events or congestion.
Similar to our work, they observe non-consecutive packet losses due to single BGP events, albeit in a different setup (inter-AS valley-free routing).
Our work extends the understanding of transient violations to diverse events and configurations, through a controllable and repeatable environment.

\paragraph{BGP convergence time on specific setups}
Many papers focus on the BGP convergence time.
Labovitz et al.~\cite{labovitz01theimpact} inject single-prefix withdraws at diverse ISPs and analyze the resulting eBGP convergence time ($T_{down}$) both theoretically and experimentally.
They show that the length of the longest possible backup path upper bounds $T_{down}$ and this upper bound is also observed on average in practice.
Labovitz et al.~\cite{labovitz00delayed} extend by injecting probe traffic to measure the degradation in end-to-end performance---concluding that withdraws are the longest event to recover from.
Park et al.~\cite{park11quantifying} passively observe iBGP updates in two ISPs to infer routing events and the resulting iBGP convergence times.
They find that route-reflection slightly increases the convergence time compared to full-mesh.
However, as we have seen (cf.\ \cref{sec:motivation}), the convergence time can be a poor proxy of reachability violation times, which is the focus of our work. 
Further, we conduct controlled experiments to avoid inference inaccuracies, and explore multi-prefix BGP events---showing that the processing time increases with the number of prefixes and dominates for >100-prefix events.

\paragraph{Single-router reaction times to BGP messages}
Feldmann et al.~\cite{feldmann04measuring} inject data-plane traffic and BGP updates to a router to measure its control-plane reaction times.
They find that while in general the average reaction times are below 150\,ms, they increase to several seconds for low-rate data-plane traffic and/or enough peers.
BGPSeer~\cite{schmid23predicting} additionally measures a router's data-plane reaction times to single-prefix withdrawals and builds a time-aware BGP simulator to predict transient violation times.
We extend this work by building a network-wide controllable environment that enables measuring transient violation times.

\paragraph{IGP transient violations}
A complementary line of work analyzes violations due to internal router/link failures.
Fran{\c{c}}ois et al.~\cite{francois05achieving,francois07thesis} decompose and measure the convergence time on a single router, and design a simulator to predict the convergence time in networks.
They show that the IGP convergence bottleneck is the time to update the RIB/FIB.
Our work complements this analysis by showing that for BGP convergence, the bottleneck is responding to the iBGP neighbors, with the FIB update completing soon after.
DCART~\cite{merindol18afinegrained} monitors an ISP's data- (internal probe traffic), control- (IS-IS routing updates), and management-plane to detect packet loss and forwarding loops.
By correlating these data sources, DCART attributes most of the long periods of loss to IGP convergence, instead of congestion/DDoS.

Other work measures transient violations without identifying the triggering event.
Hengartner et al.~\cite{hengartner02detection} measure transient loop durations in a Tier-1 ISP. 
However, the measurements are passive, which introduces inaccuracies when there are no packets to loop, and are limited to a fixed network.

\paragraph{IGP convergence time}
There is complementary work on intra-domain protocols, including comparing centralized SDN to OSPF~\cite{abdallah18networkconvergence}.

\paragraph{Preventing transient violations}
To prevent transient violations due to \emph{planned} BGP reconfigurations, Chameleon~\cite{schneider23taming} controls the ordering in which routers learn BGP routes. 
As such, it cannot prevent violations due to unforeseen external BGP events.
Our work analyzes the effect of those violations.
In the same spirit, Fran{\c{c}}ois and Bonaventure~\cite{francois07avoiding} avoid transient loops during IGP convergence but only for planned link changes or unplanned failures of links protected by Fast Reroute.

\paragraph{Stable-state verification}
State-of-the-art verifiers take as input a network configuration, a forwarding property, and a (probabilistic) failure model, and analyze whether any stable state violates the property~\cite{beckett17minesweeper}, or the likelihood of the property holding~\cite{steffen20netdice}.
While useful, all of these tools consider only the resulting stable state after the reconvergence has finished.
However, as we have shown in this work, there can be substantial transient violations of reachability despite the previous and final stable states being fully correct.

\section{Conclusion \& Future Work}

We conclude with a brief summary and an outline how our work motivates further research: i)~obtaining further insights using our measurement framework, ii)~training models to accurately predict transient violation times using our collected dataset, and iii)~reducing violation times using insights from our analysis.

\paragraph{Summary}
We have seen that transient violations resulting from a single event can persist for several minutes in extreme cases.
Multiple-prefix withdrawals, as anticipated, lead to the longest violation times.
Notably, announce and update-worse events, although smaller in scale, can still trigger reachability violations in the network due to transient forwarding loops or reverse-path filtering.

Further, we observed that long processing times pair and accumulate with the information dissemination in the network required to resolve an event, as the latter typically causes devices in the network to \emph{repeat} the processing of the event multiple times on the critical path toward resolving a violation.

Interestingly, we discovered that maintaining advertised backup routes impacts violation times in opposite ways depending on the BGP event type.
While withdraw events generally benefit from pre-disseminated backup routes throughout the network, update-worse events may actually cause shorter violations when the backup route is announced only in response to the event.

Our findings also indicate that the violation time at a router correlates with its ``total propagation delay'', meaning that routers in closer proximity to \emph{both} the original and final egresses recover faster.

\paragraph{Further insights about transient violations}
Although we focus on reachability and several network scenarios, the design of our measurement framework and open-source orchestration code lower the barrier to further analysis of transient violations.
For example, one can use our methodology to accurately measure transient violation times of any path-based forwarding property, under more involved route-maps, or due to interactions between devices of different vendors.

\paragraph{Better models of transient violations}
We envision that our measurement framework will serve as a training ground for models of transient violations.
For example, one can use our already collected datasets to train the timing model of BGPSeer~\cite{schmid23predicting} to more accurately predict violation times for multi-prefix events and route-reflection configurations.

\paragraph{Shorter transient violations}
Our measurement framework and insights can help reduce transient violation times or even entirely avoid certain violations.
First, thanks to the network-wide network observability and high-level of control of our measurement framework, one can replicate on it intricate scenarios observed ``in the wild'' to find their root causes and try to mitigate them.

Second, we observe that there are two categories of events: those with fundamentally unavoidable violations (withdraw) or not (update with a worse route/announce of a new best route).
Especially for the latter category, convergence speed might not be all we should care about: since at least one route is available at all times, we can afford to extend the convergence time to avoid reachability violations alltogether.
For example, we imagine invoking reconfiguration tools like Chameleon~\cite{schneider23taming} and Soup/Loup~\cite{gvozdiev13loup} upon an UpdateWorse event.

Third, we observe that for multi-prefix events, routers often inform their peers at different times for the same prefix---causing a ``de-synchronization'' of different routers and forwarding loops.
To avoid this, it would be interesting for a router to synchronize the sending of updates for the same prefix to its neighbors.

\subsection*{Acknowledgments}
The research leading to these results was supported by an ERC Starting Grant (SyNET) 851809.

\bibliographystyle{plain} 
\bibliography{refs}

\newpage

\appendix

\section{Emulating background traffic}
\label{appendix:background-traffic}

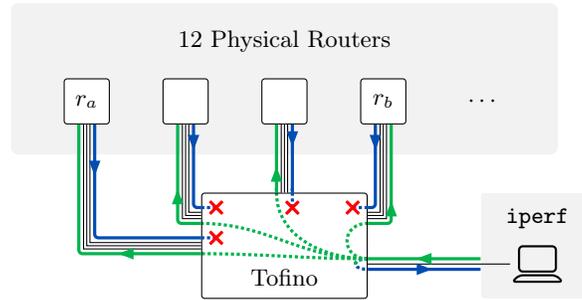
\begin{figure}[t]
  \centering
  \let\ShowProberTraffic\undefined
  \def\ShowBackgroundTraffic{1}
  \colorlet{Background}{green!70!blue}
\colorlet{BackgroundRet}{green!30!blue}
\begin{tikzpicture} [
    router/.style={rounded corners=0.4mm, draw, fill=white, minimum size=6mm},
    triple/.style={draw, preaction={draw, double, double distance=2pt}},
    box/.style={draw, rounded corners=0.4mm},
    component/.style={box, draw=none, fill=black!5, inner sep=0pt, minimum height=14mm},
    database/.style={cylinder, draw, shape border rotate=90, aspect=0.5, path picture={
        \coordinate (w1) at ($(path picture bounding box.north west)!0.3!(path picture bounding box.south west)$);
        \coordinate (e1) at ($(path picture bounding box.north east)!0.3!(path picture bounding box.south east)$);
        \coordinate (w2) at ($(path picture bounding box.north west)!0.5!(path picture bounding box.south west)$);
        \coordinate (e2) at ($(path picture bounding box.north east)!0.5!(path picture bounding box.south east)$);
        \coordinate (w3) at ($(path picture bounding box.north west)!0.7!(path picture bounding box.south west)$);
        \coordinate (e3) at ($(path picture bounding box.north east)!0.7!(path picture bounding box.south east)$);
        \draw (w1) to[out=-90, in=-90, looseness=0.5] (e1);
        \draw (w2) to[out=-90, in=-90, looseness=0.5] (e2);
        \draw (w3) to[out=-90, in=-90, looseness=0.5] (e3);
    }},
    clock/.style={circle, draw, path picture={
      \coordinate (c) at (path picture bounding box.center);
      \draw[fill] (c) circle (0.2mm);
      \draw (c) --++ (150:1.7mm);
      \draw (c) --++ (30:2.3mm);
    }},
    laptop/.style={path picture={
      \coordinate[yshift=-0.4pt] (screen nw) at ($(path picture bounding box.north west)!0.15!(path picture bounding box.north east)$);
      \coordinate (screen height w) at ($(path picture bounding box.north west)!0.7!(path picture bounding box.south west)$);
      \coordinate (screen height e) at ($(path picture bounding box.north east)!0.7!(path picture bounding box.south east)$);
      \coordinate (screen se) at ($(screen height w)!0.85!(screen height e)$);
      \coordinate[yshift=-0.5mm] (keys nw) at ($(screen nw |- screen se)$);
      \coordinate[yshift=-0.5mm] (keys ne) at (screen se);
      \coordinate[yshift=+0.4pt, xshift=+0.4pt] (keys sw) at (path picture bounding box.south west);
      \coordinate[yshift=+0.4pt, xshift=-0.4pt] (keys se) at (path picture bounding box.south east);
      \draw[rounded corners=0.5mm] (screen nw) rectangle (screen se);
      \draw[rounded corners=0.5mm] ($(keys nw)!0.5!(keys ne)$) -- (keys nw) -- (keys sw) -- (keys se) -- (keys ne) -- cycle;
    }},
    drop/.style={minimum width=2mm, minimum height=2mm, inner sep=0pt, path picture={
      \draw[very thick, draw=red, solid] (path picture bounding box.north east) -- (path picture bounding box.south west);
      \draw[very thick, draw=red, solid] (path picture bounding box.north west) -- (path picture bounding box.south east);
    }},
    mid arrow/.style={postaction={decorate,decoration={
          markings,
          mark=at position #1 with {\arrow{Triangle[length=6pt, width=4pt, sep=-3.2pt]}}
    }}},
    iface/.style={coordinate},
    bg traffic/.style={very thick, Background, rounded corners=0.5mm},
    bg return/.style={bg traffic, BackgroundRet},
  ]

  \node[router] (r1) {$r_{a}$};
  \node[router, right=7mm of r1] (r2) {};
  \node[router, right=7mm of r2] (r3) {};
  \node[router, right=7mm of r3] (r4) {$r_{b}$};
  \node[router, right=7mm of r4, opacity=0, label=center:$\cdots$] (rn) {};
  \begin{pgfonlayer}{background}
    \node[rounded corners=1mm, fit={(r1) (r2) (r3) (r4) (rn)}, draw=none, fill=black!5, inner sep=7mm, yshift=3mm] (physical) {};
  \end{pgfonlayer}
  \node[yshift=5mm] at (physical) {12 Physical Routers};

  \node[box, minimum width=22mm, minimum height=14mm, below=5mm of physical, label={[yshift=-13.5mm]Tofino}] (tofino) {};

  \draw[triple] (r1) |- (tofino.180) node[coordinate] (tofino if1) {};
  \draw[triple] (r2) |- (tofino.160) node[coordinate] (tofino if2) {};
  \draw[triple] (r3) -- (tofino.north) node[coordinate] (tofino if3) {};
  \draw[triple] (r4) |- (tofino.20) node[coordinate] (tofino if4) {};

  \node[component, minimum width=1.5cm, right=15mm of tofino] (iperf) {};
  \node[below=1mm of iperf.north, align=center] (iperf text) {\texttt{iperf}};
  \node[laptop, thick, minimum width=7mm, minimum height=5mm, above=2mm of iperf.south] (iperf laptop) {};
  \node[iface] at (iperf.west |- iperf laptop) (iperf if) {};
  \draw[] (iperf laptop) -- (tofino.east |- iperf if) node[coordinate] (tofino if5) {};

  \draw[mid arrow=0.5, bg traffic] ([yshift=2pt]iperf if) -- ([yshift=2pt]tofino if5) node[coordinate] (bg p1) {};
  \draw[mid arrow=0.5, bg return] ([yshift=-2pt]tofino if5) node[coordinate] (ret p5) {} -- ([yshift=-2pt]iperf if);

  \draw[mid arrow=0.3, bg traffic] ([yshift=-3pt]tofino if1) node[coordinate] (bg p3) {} -| ([xshift=-3pt]r1.south);
  \draw[mid arrow=0.4, bg traffic] ([yshift=-3pt]tofino if2) node[coordinate] (bg p4) {} -| ([xshift=-3pt]r2.south);
  \draw[mid arrow=0.3, bg traffic] ([xshift=-3pt]tofino if3) node[coordinate] (bg p5) {} -| ([xshift=-3pt]r3.south);
  \draw[mid arrow=0.3, bg traffic] ([yshift=-3pt]tofino if4) node[coordinate] (bg p6) {} -| ([xshift=+3pt]r4.south);

  \draw[mid arrow=0.2, bg return] ([xshift=+3pt]r1.south) |- ([yshift=3pt]tofino if1) node[coordinate] (ret p1) {};
  \draw[mid arrow=0.2, bg return] ([xshift=+3pt]r2.south) |- ([yshift=3pt]tofino if2) node[coordinate] (ret p2) {};
  \draw[mid arrow=0.2, bg return] ([xshift=+3pt]r3.south) |- ([xshift=3pt]tofino if3) node[coordinate] (ret p3) {};
  \draw[mid arrow=0.2, bg return] ([xshift=-3pt]r4.south) |- ([yshift=3pt]tofino if4) node[coordinate] (ret p4) {};

  \draw[bg return, densely dotted] (ret p1) -- ++(2mm, 0) node[drop] {};
  \draw[bg return, densely dotted] (ret p2) -- ++(2mm, 0) node[drop] {};
  \draw[bg return, densely dotted] (ret p3) -- ++(0,-2mm) node[drop] {};
  \draw[bg return, densely dotted] (ret p4) -- ++(-2mm,0) node[drop] {};

  \begin{pgfonlayer}{background}
    \draw[bg traffic, densely dotted] (bg p1) to[out=180, in=0] (bg p3);
    \draw[bg traffic, densely dotted] (bg p1) to[out=180, in=0] (bg p4);
    \draw[bg traffic, densely dotted] (bg p1) to[out=180, in=270] (bg p5);
    \draw[bg traffic, densely dotted] (bg p1) to[out=180, in=180, looseness=2] (bg p6);

    \draw[bg return, densely dotted] (bg p1) to[out=180, in=180, looseness=4] (ret p5);
    \begin{scope}
      \clip ([yshift=2pt]bg p1) rectangle ++(-30pt, -4.2pt);
      \draw[bg traffic, densely dotted] (bg p1) to[out=180, in=180, looseness=4] (ret p5);
    \end{scope}
  \end{pgfonlayer}

\end{tikzpicture}

  \caption{
    Background traffic from \texttt{iperf} is duplicated by the Tofino for all routers and all interfaces
  }
  \label{fig:testbed-background-traffic}
\end{figure}

In order to exclude artifacts of low network usage, we generate constant-rate data-plane traffic as background traffic on the network.
We run an \texttt{iperf3} server and client on one server (using network namespaces to separate the two processes and force the traffic through the physical network interfaces), thus generating a constant rate of UDP traffic flowing over the programmable Intel Tofino switch.
The Tofino then duplicates this traffic and changes the destinaion MAC and IP address to hit all the ingress and egress queues of all routers.
It then drops the packets being sent back from the routers to the Tofino.
This setup ensures evenly spread background traffic on each link, without requiring to construct sophisticated paths through the network that would hit each link with a precise amount of traffic.

For our experiments, we generate 2.5\,Gbps data-plane background traffic, ensuring a base link utilization of 25\% on each of the routers' 10\,Gbps links.

\end{document}